\documentclass[fleqn,usenatbib]{mnras}%

\usepackage{newtxtext,newtxmath}

\usepackage[T1]{fontenc}

\DeclareRobustCommand{\VAN}[3]{#2}
\let\VANthebibliography\thebibliography
\def\thebibliography{\DeclareRobustCommand{\VAN}[3]{##3}\VANthebibliography}

\usepackage[english]{babel}
\usepackage[utf8]{inputenc}

\usepackage{amsmath}
\usepackage[labelfont=bf]{caption}
\usepackage{fancyhdr}
\usepackage{graphicx}
\usepackage{ulem}

\usepackage{xcolor}
\usepackage{subfigure}

\definecolor{Red}{rgb}{0.9,0,0}
\definecolor{Blue}{rgb}{0,0,0.9}
\definecolor{Green}{rgb}{0,0.5,0}
\definecolor{Black}{rgb}{0,0,0}

\title[SORA: Stellar Occultation Reduction and Analysis]{SORA: Stellar Occultation Reduction and Analysis}

\author[Gomes-Júnior, A. R., Morgado, B. E. et al.]{Gomes-Júnior, A. R.$^{1,2}$\thanks{E-mail: altair.gomes@unesp.br, altairgomesjr@gmail.com},
Morgado, B. E.$^{3,2,4}$,
Benedetti-Rossi, G.$^{1,2,3}$,
Boufleur, R. C.$^{2}$,\newauthor
Rommel, F. L.$^{4,2}$,
Banda-Huarca, M. V.$^{4,2}$,
Kilic, Y.$^{5,6}$,
Braga-Ribas, F.$^{7,3,4,2}$
and Sicardy, B.$^{3}$
\\
$^1$UNESP - S\~ao Paulo State University, Grupo de Din\^amica Orbital e Planetologia, CEP 12516-410, Guaratinguet\'a, SP 12516-410, Brazil\\
$^2$Laborat\'orio Interinstitucional de e-Astronomia - LIneA and INCT do e-Universo, Rua Gal. Jos\'e Cristino 77, Rio de Janeiro, RJ 20921-400, Brazil\\
$^3$LESIA, Observatoire de Paris, Université PSL, CNRS, Sorbonne Université, Univ. Paris Diderot, Sorbonne Paris Cité, 5 place Jules Janssen, 92195 Meudon, France\\
$^4$Observat\'orio Nacional/MCTIC, R. General Jos\'e Cristino 77, Rio de Janeiro, RJ 20.921-400, Brazil\\
$^5$Akdeniz University, Faculty of Sciences, Department of Space Sciences and Technologies, 07058 Antalya, Turkey\\
$^6$TÜB\.{I}TAK National Observatory, Akdeniz University Campus, 07058 Antalya, Turkey\\
$^7$Federal University of Technology-Paraná (UTFPR / DAFIS), Curitiba, Brazil
}

\date{Accepted XXX. Received YYY; in original form ZZZ}

\pubyear{2021}

\begin{document}
\label{firstpage}
\pagerange{\pageref{firstpage}--\pageref{lastpage}}
\maketitle

\begin{abstract}

The stellar occultation technique provides competitive accuracy in determining the sizes, shapes, astrometry, etc., of the occulting body, comparable to in-situ observations by spacecraft. With the increase in the number of known Solar System objects expected from the LSST, the highly precise astrometric catalogues, such as Gaia, and the improvement of ephemerides, occultations observations will become more common with a higher number of chords in each observation. In the context of the Big Data era, we developed \textsc{sora}, an open-source python library to reduce and analyse stellar occultation data efficiently. It includes routines from predicting such events up to the determination of Solar System bodies' sizes, shapes, and positions.

\end{abstract}

\begin{keywords}
Occultations -- Methods: data analysis -- Methods: miscellaneous
\end{keywords}

\section{Introduction}\label{sec:introduction}

A stellar occultation is an event that happens when an occulting body (hereafter just referred as body) comes into an observer's line of sight with respect to a star. At that moment, the body's shadow may be observed as the drop in the total stellar light flux measured. The observation can be converted into a path on the sky, called \textit{chord}, given the relative velocity of the observer with respect to the body. The chord's accuracy depends primarily on the accuracy of the times obtained from the light curve.

When many sites observe the same occultation, each will obtain chords of different sizes if they are separated perpendicular to the movement of the shadow. Considering all these chords, it is possible to determine a limb profile and calculate the body's size. It is important to note that, unlike an eclipse, the star is farther and its angular size is much smaller than the body's, which results in almost the same size for the body and its umbra. Because of this, the stellar occultation technique can easily provide us with the bi-dimensional apparent size and shape of the Solar System body with high accuracy, in many cases without the need to know much about the star beside its position. The accuracy achieved in determining the sizes of Solar System bodies from ground-based stellar occultations is only surpassed by in-situ space-based observations.

The advantage of the stellar occultation technique is not limited to the precise measurement of the occulting body's size. The centre of the body is obtained relative to the star, meaning that the final astrometric position is limited by the uncertainty on the position of the star \citep{Rommel2020}. Topographic features as craters, mountains, chasms, among others, can also be detected from the limb profiles or variations in the light curve, as shown in \cite{DiasOliveira2017}. In \cite{GomesJunior2019}, they improved the determination of the rotational period of the Saturnian irregular satellite Phoebe by comparing a 3D shape model to the observed occultation chords.

The technique can also provide information about the body's environment and surroundings. For instance, the atmosphere of Pluto was discovered by a stellar occultation given the smooth disappearance and reappearance of the star caused by the refraction of the light on the atmosphere \citep{Millis1993,Meza2019}. In addition, four out of six ring systems known to date in planets and small bodies in the Solar System were also discovered using this technique: Uranus \citep{Elliot1977}; Neptune \citep{Sicardy1991}; Chariklo \citep{BragaRibas2014}; and Haumea \citep{Ortiz2017}.

Furthermore, occultations can also reveal some physical parameters of the star. A system with multiple stars may cause multiple occultations, and more than one drop can be observed in the light curve \citep{Berard2017,Leiva2020}. If the star's angular size projected at the distance of the body is large enough, the penumbra may be noticeable as a slower disappearance and reappearance \citep{Widemann2009,Levine2021}.

On the other hand, stellar occultations by small bodies have always been difficult to predict and observe. A reliable prediction demands the accurate position of the occulted star and occulting body ephemeris. Considering that the umbra has nearly the same size as the occulting body, its position needs to be known with an accuracy of the order of the occulting body's angular size. For instance, to ensure an occultation observation by Ceres, the required precision of the prediction must be better than 230 milliarcseconds (mas), while for Pluto, an accuracy of 40 mas is needed. Although feasible, this level of precision has been easily achieved recently with the publication of more precise astrometric catalogues, which improved the position of the stars themselves and allowed better astrometric measurements of the Solar System bodies and, in consequence, better ephemeris.

With the release of the Gaia catalogue, the position of the stars is now very accurate. For instance, the Gaia-EDR3 \citep{Brown2021} catalogue has positions for more than 1.8 billion sources with uncertainties below 1 mas for objects as faint as magnitude $G=21$. For stars brighter than $G=17$, the uncertainty of their positions can be smaller than 50 $\mu$as. These Gaia catalogues allowed accurate predictions of stellar occultations leading to an increase in the number of events observed over the years \citep{BragaRibas2019, Herald2020}.

Moreover, the Legacy Survey of Space and Time (LSST) at the Vera C. Rubin Observatory, expected to start operating in 2024 for about ten years, will provide many scientific outcomes, including positions of Solar System bodies and an unprecedented number of discoveries. It is expected that the LSST will observe more than 30.000 TNOs with $R < 24.5$ \citep{Ivezic2019}, with the majority discovered by the survey.

The combination of those factors will allow that the number of reliable predicted and observed stellar occultations to significantly increase \citep{Camargo2018}. To better benefit from this new Big Data era, we have developed a suitable reduction procedure that can be implemented in a more efficient, precise, and faster pipeline. 

Stellar occultations have been observed for many years, and there are a few software available that analyse these events — each one of them with a specific purpose and capabilities. For instance, the \textsc{occular}, \textsc{rote}, and \textsc{pyote}\footnote{\url{http://occultations.org/observing/software/ote/}} software can extract the occultation timings from the light curves, and \textsc{Occult}\footnote{\url{http://www.lunar-occultations.com/iota/occult4.html}} is able to make predictions of stellar occultations and analyse the event.

The here presented open-source Python library \textsc{sora} (Stellar Occultation Reduction and Analysis) contains methods for the analysis of stellar occultations. \textsc{sora} is based on Python classes with tools to handle the star, body, observer, and light curve information that can be integrated into a dynamical reduction methodology of data generated by occultation events. The nature of the library requires the user to have previous knowledge of the python language and a good understanding of the occultation process to choose appropriate parameters. Most of the parameters in \textsc{sora} can be changed, thus allowing a more personalised reduction.

In \autoref{sec:occultation} we describe the stellar occultation technique: how to make predictions; the important features in the light curve to determine the instants of immersion and emersion; how to project the occultation chords on the tangent plane; and possible results obtained from this technique. It is an overview of the science and the reduction procedure present in \textsc{sora} without getting into details of the particularities of the software. In \autoref{sec:sora} we describe the \textsc{sora} library itself: how \textsc{sora} works; the tasks currently implemented; the fitting procedures; and its capabilities. Finally, in \autoref{sec:conclusion} we summarise and discuss about the future of \textsc{sora}.

\section{Stellar occultation Technique}\label{sec:occultation}

A stellar occultation can provide many physical characteristics of the occulting body, such as size and profile, with the quality compared to in-situ observations made by spacecraft. This is possible by indirectly observing the body through its apparent interaction with the star. Accurate coordinates are known for the star from catalogues such as Gaia. Thus, it is possible to associate the instants of the star's disappearance (immersion) and reappearance (emersion) with the object's position in the sky, converting the time resolution of the observations into spatial resolution.

To determine the instants of the occultation, we should be able to identify the magnitude drop in the light curve. With a higher Signal-to-Noise Ratio (S/N), we can identify even small drops, like the ones caused by rings or occultation involving bodies brighter than the occulting star, like the occultations by the Galilean satellites \citep{Morgado2019}. S/N is proportional to the square root of the exposure time, thus increasing the exposure, we obtain higher S/N. At the same time, with higher exposure, we may miss the details of the light curve like the size of the star and diffraction pattern, or, for events with a short duration, the occultation may be embedded in one exposure. Thus, it is essential to have a short exposure time, which decreases the S/N. In conclusion, the observations must present a suitable S/N, considering an exposure time as short as possible to optimise the detection of the event.

The success of a stellar occultation depends on a few factors: predictions with low uncertainties to optimise the areas where observations must be taken place (\autoref{Subsec:science-prediction}); a light curve from the observations with the highest S/N and time resolution as possible where sub-exposure times of immersion and emersion can be determined (\autoref{Subsec:science-lightcurve}); the time and position of each observer must be synchronised with a reliable time source, for instance using GPS so that all observers can be referred to the same reference system (\autoref{Subsec:science-geometry}); and finally, precise reduction procedure that considers all the concerning effects such as diffraction, Earth's precession and nutation, and combines all the observations to determine the physical parameters of the body (\autoref{Subsec:science-results}).

\subsection{Predicting stellar occultations}
\label{Subsec:science-prediction}

Observing a positive stellar occultation requires a specific geometry where the Solar System body comes into an observer's line of sight with respect to a star. Unless the observer can freely move into this circumstance, we must wait for that to happen. However, if the body moves in front of a high star density region in the sky plane, the chances of an occultation increase \citep{GomesJunior2016}.

To predict when and where such an event can be observed, the combined positions of the stars and the ephemeris must present uncertainties of the order of the body's apparent radius, as stated before. For a closer and large body like Ceres, it is about 230 mas, while for farther and smaller bodies, it can be as low as 1 mas. Thus, it is essential to use the best star catalogues and ephemeris available.

With the current version of \textsc{sora}, the positions of the stars can be retrieved from Gaia-DR2 \citep{Brown2018} or Gaia-EDR3 \citep{Brown2021} catalogues. With Gaia-EDR3, the position of the stars is known with precisions better than 1 mas, reaching 10 $\mu$as for $G<15$ stars at the catalogue epoch. Due to very precise proper motions ($\sim 20\ \mu \textrm{as} ~ \textrm{yr}^{-1}$ for $G<15$), Gaia stars may still maintain high precision when propagating the position farther from the catalogue epoch. To make use of the highly precise Gaia proper motions, the rigorous stellar motion described by \cite{Butkevich2014} is used.

Even so, some Gaia-EDR3 positions must be handled with care. For instance, the \textit{renormalised unit weight error} (RUWE) parameter tells us how much the centre of the star deviates from the standard model of stellar motion \citep{Lindegren2021}. This parameter has been associated with unresolved binary stars \citep{Stassun2021} which may affect the astrometric position estimated at the occultation epoch. \cite{Lindegren2018} proposed a $\textrm{RUWE} <= 1.4$ as a criterion for good solutions. Nevertheless, positive occultations with $\textrm{RUWE} > 1.4$ were observed and showed a significant shift from the predictions \citep{Dunham2021JOA}.

Furthermore, \cite{CantatGaudin2021} found a systematic error up to $80\ \mu \textrm{as} ~ \textrm{yr}^{-1}$ in the proper motion of the Gaia-EDR3 catalogue for $G<13$ stars, a value higher than the formal uncertainties. As an example, the star occulted by Pluto on June $09^{th}$, 1988 \citep{Millis1993} presents a significant difference in position given the corrected and uncorrected proper motions of $\Delta\alpha\cos\delta=0.08\ \textrm{mas}$ and $\Delta\delta = 1.10\ \textrm{mas}$ while the formal uncertainty at occultation epoch is $\sigma\alpha = 0.45\ \textrm{mas}$ and $\sigma\delta = 0.35\ \textrm{mas}$.

On the other hand, the ephemeris used must also present high accuracy. For that, the astrometric positions (for old and new observations) used in orbital fitting must be referred to the ICRS using the best catalogue available. The re-reduction of old observations with Gaia is well encouraged \citep{Arlot2012}.

Since occultation predictions are focused on the near future events, dedicated ephemeris, such as those generated by NIMA \citep{Desmars2015}, is of fundamental importance. The ability to update and provide ephemerides as new observations are furnished is crucial to minimise the uncertainty of near events. Furthermore, the observations should be adequately weighted based on the catalogue used for reduction, the number of positions in the night, etc. \citep{Desmars2015}. Using the astrometric position resulting from a stellar occultation itself, whose position mainly depends on the position of the occulted star, can highly improve the ephemeris \citep{Desmars2019}, even if it is from a single-chord event \citep{Rommel2020}.

Alternatively, an ephemeris offset of the body a few days before the occultation can be determined with new observations and used to correct the ephemeris and improve the prediction. This method was adopted with great success by \cite{Assafin2012, Camargo2013} and \cite{Ortiz2020b}, to mention a few works. The technique assumes the offset is constant over a few days, which may be good enough in the majority of the cases. However, the observation and astrometric reduction provide positions with uncertainties in tenths to hundreds of mas, which may be larger than the body's angular size.

Finally, the prediction can be made by the direct cross-match between the ephemeris and star position at epoch. The distance of the apparent closest approach for a given observer must be smaller than the angular radius of the body for the occultation to be detected. Thus, if known, the uncertainty of the ephemeris can be used to increase the range of search for potential events. For occultations on Earth, the ephemeris and the star position can be geocentric, instead of topocentric or referred to a spacecraft, and the Earth's radius is added to the range of search. With this, it is possible to obtain all occultations whose shadows pass on Earth.

\begin{figure}
    \centering
    \includegraphics[width=\linewidth]{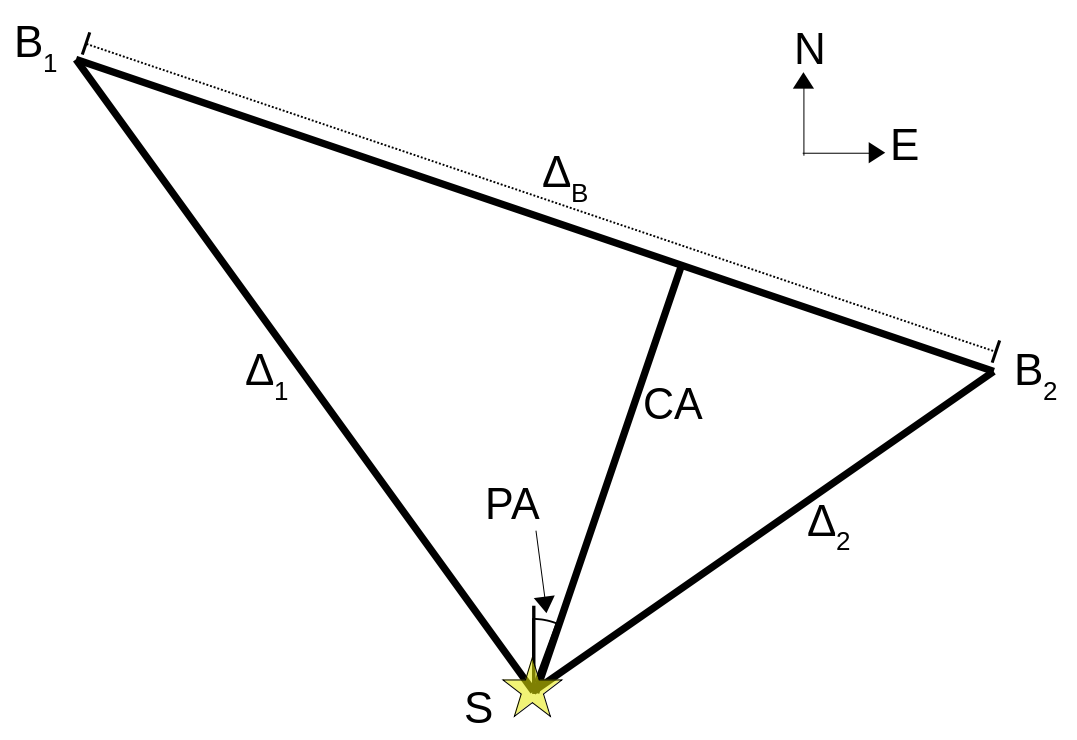}
    \caption{Representation of an occultation geometry. $S$ is the position of the candidate star, $B_1$ and $B_2$ are the body ephemerides at times $t_1$ and $t_2$, respectively, and closest to the star. $\Delta_1$ and $\Delta_2$ are the star angular separation to $B_1$ and $B_2$, respectively, while $\Delta_B$ is the angular separation between the ephemerides. $CA$ is the distance of closest approach. PA is the position angle of the body's closest approach to the star. N and E give the celestial North and East direction on the tangent plane.}
    \label{fig:occ_geometry}
\end{figure}

The duration of a stellar occultation will depend on the velocity of the occulting body relative to the star and the observer, usually less than a few minutes. For example, considering only the Earth's mean orbital velocity of $29.78\ \textrm{km} ~ \textrm{s}^{-1}$, the small body's shadow would cross the Earth in about 7 minutes. Thus, we can assume a linear shadow movement relative to the observer during the event. Therefore, we calculate the occultation parameters: distance and instant of closest approach (CA); position angle (PA); and shadow velocity ($V_S$). For that, we use the geometry shown in \cite{Assafin2010} and reproduced in \autoref{fig:occ_geometry}.

For two consecutive ephemeris positions ($B_1$, $B_2$, associated with respective times $t_1$ and $t_2$, where $t_2>t_1$) close to the star ($S$), the distance ($CA$) and instant ($t_0$) of closest approach are determined using equations \ref{eq:CA} and \ref{eq:tCA}, respectively:

\begin{equation}\label{eq:CA}
    CA = \sqrt{\Delta_1^2 - \left(\frac{\Delta_1^2 - \Delta_2^2 + \Delta_B^2}{2\Delta_B} \right)^2},
\end{equation}

\begin{equation}\label{eq:tCA}
    t_0 = t_1 + (t_2 - t_1)\sqrt{\frac{\Delta_1^2 - (CA)^2}{\Delta_B^2}},
\end{equation}
where $\Delta_1$ and $\Delta_2$ are the respective angular separation of the star to $B_1$ and $B_2$, and $\Delta_B$ is the angular separation between $B_1$ and $B_2$.

The shadow velocity ($V_S$) at $t_0$ is determined using:
\begin{equation}
    V_S = \frac{D_\circ\sin(\Delta_B)}{t_2-t_1},
\end{equation}
where $D_\circ$ is the radial distance to the occulting body. Finally, the position angle can be easily calculated from the ephemeris at $t_0$ and the star positions using standard procedures. Moreover, these parameters can be used to draw an occultation map.

It is important to note that this procedure may not provide an accurate shadow path if the shadow's velocity is too low, for instance, when the object is in quadrature. A polynomial may be used to represent the difference between the ephemeris and star positions in such cases. Usually, this is not a problem since the difference from a linear path may be much smaller than the ephemeris uncertainty, for less precise small bodies, or the shadow size, for large bodies.

\subsection{Light Curve analysis}
\label{Subsec:science-lightcurve}

The light curve will present a drop (or multiple drops) caused by the occultation in a positive detection. Different characteristics in the light curve will reveal other physical characteristics of the body and its environment, such as atmospheres \citep{Sicardy2016,Meza2019}, rings \citep{BragaRibas2014,Ortiz2017}, topographical features \citep{DiasOliveira2017}, the presence of satellites or binarity \citep{Berard2017, Leiva2020}. Intrinsic characteristics of the occulted star (or star system) may also be translated into features present in the light curve. For example, if the angular size is significant, it might generate a penumbra \citep{Widemann2009,Levine2021}; multiple systems can introduce multiple flux drops \citep{Leiva2020}; stellar variability may introduce systematic trends. The current version of \textsc{sora} is solely focused on the analysis of occultations by bodies without atmosphere by a single star (see \autoref{sec:conclusion}). Because of this, here we will discuss only occultations in these scenarios.

From the images, it is impossible to separate the flux from the star and the body when the distance between them is smaller than the Point Spread Function (PSF). We call the total flux measured (star +  body) as the \textit{baseflux} of the light curve. When the star is occulted, its light may be blocked entirely, and only the light reflected by the body is measured (or the brightness of the sky if the body is too faint), which we call the \textit{bottomflux}. Given the magnitudes of the star ($M_\star$) and the body ($M_\circ$) in a given band, the magnitude drop ($\Delta M$) during an occultation can be estimated using \autoref{Eq:mag-drop}.

\begin{equation} \label{Eq:mag-drop}
    \Delta M = M_\circ - M_\star + 2.5\log(1+10^{0.4(M_\star - M_\circ)}) ~.
\end{equation}

\autoref{Eq:mag-drop} must be used to estimate the magnitude drop, and consequently, to calculate the minimum SNR required for the magnitude drop to be detected. $M_\circ$ depends on the rotational phase, which usually is unknown given the typical high uncertainties in the rotational period estimation. $M_\circ$ and $M_\star$ also depend on the instrument setup used by the observer (filter, camera). Thus, different magnitude drops are expected to be detected by different observers. In the light curve analysis, the \textit{bottomflux} must be calculated independently for each observer.

\begin{figure*}
\center
\subfigure[diffraction_wavefront][Simulated 2D diffraction pattern]{\includegraphics[width=.49\linewidth]{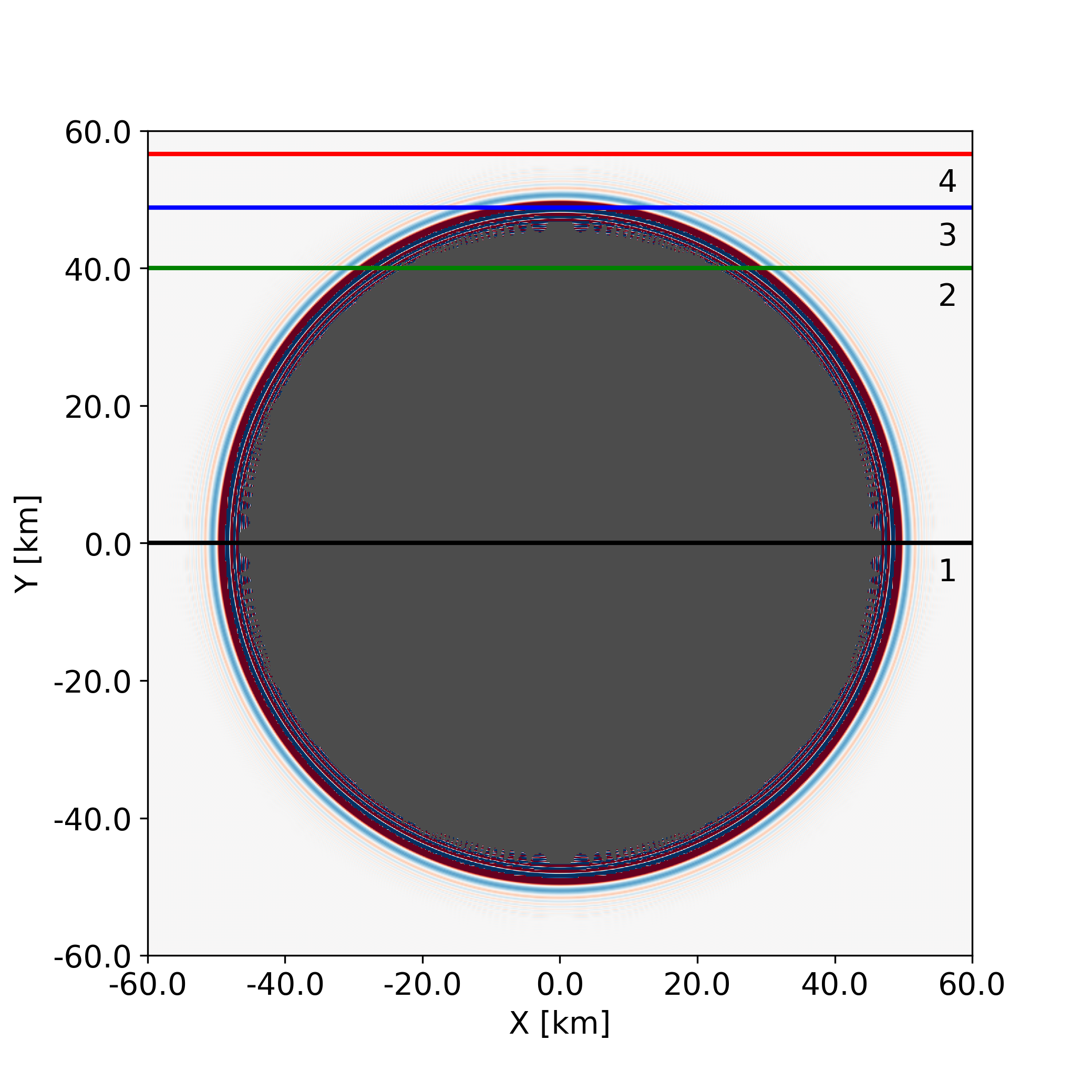} \label{fig:diffraction_wavefront}
}
\subfigure[diffraction_curve][Simulated light curves]{\includegraphics[width=.49\linewidth]{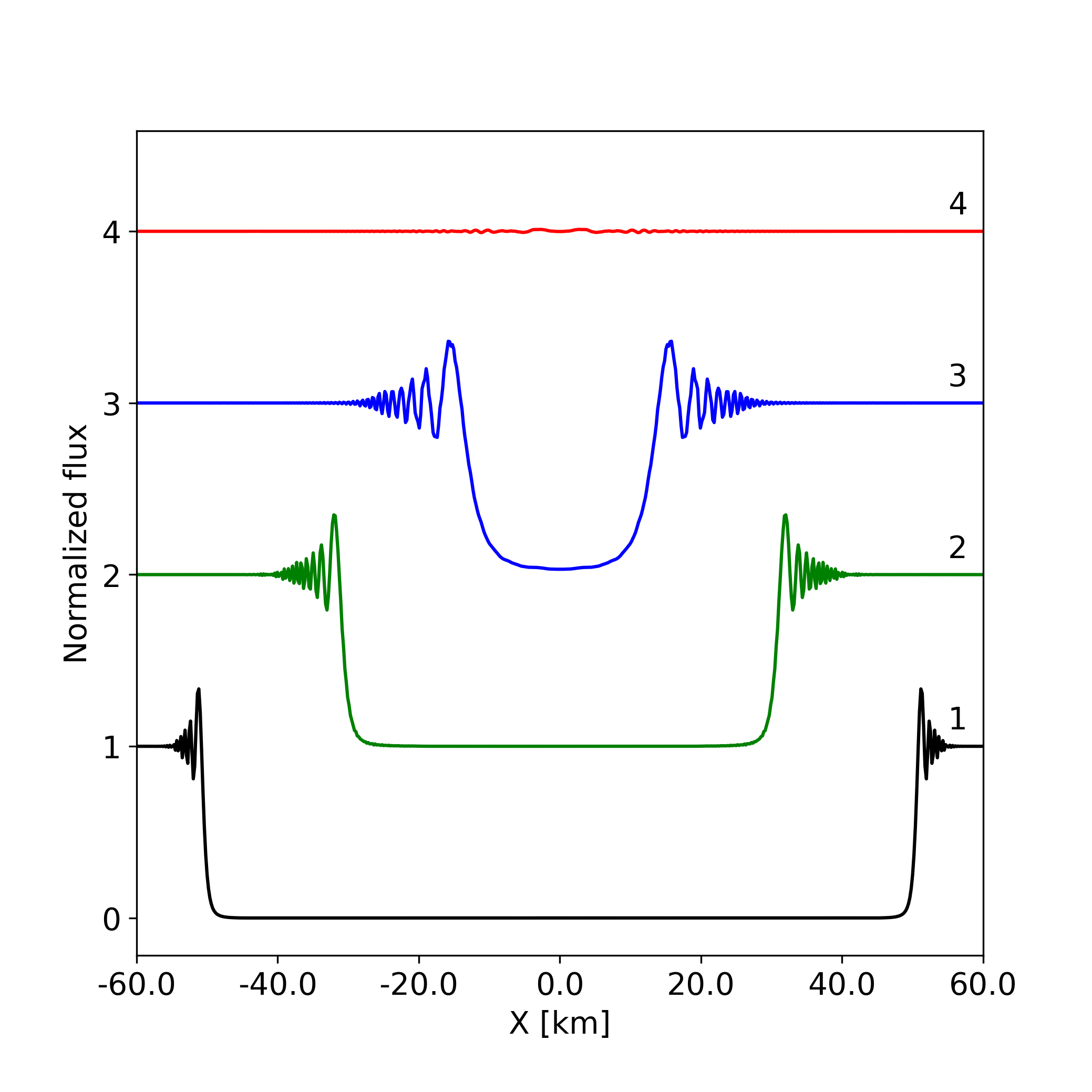} \label{fig:diffraction_curve}
}
\caption{Simulated Fresnel diffraction of a wavefront with wavelength $\lambda = 700\ \textrm{nm}$ caused by a circular body with radius $R_\circ = 50\ \textrm{km}$ and a distance from the observer $D_\circ = 20\ \textrm{au}$. The Fresnel scale is $f = 1.02\ \textrm{km}$. (a) shows the 2D diffraction pattern generated with \textsc{poppy}, an open-source optical propagation Python package originally developed for the James Webb Space Telescope project \citep{Perrin2012}. Each horizontal line represents the relative movement of the star behind the body as seen by an observer, i.e., a chord. (b) shows the light curves observed by the four chords in (a). (1) is a central chord; (2) is an intermediary chord; (3) is a grazing chord; (4) is a negative chord. \label{fig:diffraction}}
\end{figure*}

Furthermore, as described by \cite{Roques1987}, light diffraction must be taken into account. For example, \autoref{fig:diffraction_wavefront} shows the diffraction effects of a wavefront caused by an opaque 2D body with sharp edges (the so-called Fresnel diffraction). The diffraction will change the apparent size of the shadow and make the light curve in the immersion and emersion more smooth, as can be seen in \autoref{fig:diffraction_curve}.

The broadening of the shadow caused by the diffraction observed in the wavelength $\lambda$ from a distance to body $D_\circ$ can be estimated by the Fresnel scale $f=\sqrt{\lambda D_\circ / 2}$, measured in length. For a typical observation in the visible light ($\lambda \sim 550\ \textrm{nm}$), the Fresnel scale for a Plutino ($D_\circ \sim 39.5\ \textrm{au}$), for example, is $f \sim 1.2\ \textrm{km}$.

\cite{Roques1987} showed the diffraction pattern for a variety of body shapes. In these cases, they use bodies between $1$ and $10\ \textrm{km}$ in size at the distance of Uranus. The profiles broadly differ depending on the shape. However, considering that observing occultations by such small bodies at such distances is far from current detections, we expect the Fresnel scale to be much smaller than the body itself. Because of this, we can restrain its analysis to the immersion and emersion times.

Since the light curve from the observations is a path through the body, as shown in \autoref{fig:diffraction}, and the real shape is usually unknown, we consider, as a first approximation, only a one-dimensional diffraction model for the light curve. Hence, we approximate the diffraction model using the bar shape from \cite{Roques1987}. Furthermore, since observations are usually not monochromatic, the model should be integrated over the observed bandwidth.

Another important physical parameter that may affect the light curve is the size of the star projected at the distance of the occulting body. If significant, the penumbra may be detected in the light curve as the star slowly disappears behind the body. An example was shown by \cite{Widemann2009} in an occultation of a Hipparcos star by the Uranian satellite Titania. With an angular diameter of $0.54\ \textrm{mas}$, the star's apparent size projected at Titania's distance was $7.5\ \textrm{km}$. More recently, from an occultation by the Plutino (28978) Ixion, a diameter of $\sim20$ km at the body's distance was measured for the occulted star \citep{Levine2021}.

Gaia-EDR3 does not provide the stellar radius, but this parameter can be found in Gaia-DR2 \citep{Andrae2018} for 77 million stars. Given the distance of the star, the apparent size at the distance of the body can be easily calculated from trigonometry: $r_\star = D_\circ \cdot R_\star / D_\star$, where $r_\star$ and $R_\star$ are the star's apparent and physical radius, respectively, and $D_\star$ is the distance to the star. However, it is important to note that the Gaia-DR2 presents an error in its parallax estimation, and the correction provided by \cite{BailerJones2018} is encouraged to be used instead.

For stars whose sizes are not known, an alternative method is to estimate their sizes from empirical models such as those provided by \cite{vanBelle1999} and \cite{Kervella2004}, for instance. Both models use the stars' B-, V- and K-broadband photometry to predict their angular sizes. In \cite{vanBelle1999}, they provide a model with three different sets of parameters, depending on if the star is a main sequence, giant, or supergiant. \cite{Kervella2004}'s model, however, is valid for dwarf stars and subgiants.

In the light curve model previously described, the size of the star can be added by integrating the diffraction model over the area of the angular size of the star. This is correct because we can interpret the star as a collection of point sources, where each one will generate a diffraction pattern as previously described.

We may notice that if the size of the star or diffraction is important in the light curve, their effects will be seen differently depending on where the chord is located relative to the centre of the body. This is clear in \autoref{fig:diffraction_curve} where the diffraction model presents a more prominent scale far from the centre path. Grazing occultations may not present a \textit{bottomflux} as determined by \autoref{Eq:mag-drop} if the star is partially occulted.

Considering a chord that is almost tangent to the object's limb, the stellar occultation will be slower since the velocity of the stellar ingress/egress depends on the relative stellar motion direction with respect to the local limb. In these cases, the model of the light curve with the bar shape can be adapted using the velocity of the occultation normal to the surface $V_N = \Vec{V}_S \cdot \Vec{N}$, where $\Vec{N}$ is the unit normal vector.

Finally, observations are made with a given exposure time ($t_{exp}$). Each observer must choose this exposure to maximise the S/N but be short enough to have many data points during the occultation. It is important to remember that occultation converts time resolution into spatial resolution. The spatial representation of a exposure is simply calculated using $r_t = V_S \cdot t_{exp}$. If the exposure is long enough, it will dominate over the Fresnel diffraction ($f$) and the star's apparent size ($r_\star$), which are usually in the kilometre or sub kilometre order.

The diffraction model, convolved by the photometric band and the star size, using the velocity normal to the surface, must also be integrated by the exposure duration. Then, we can determine a light curve model and compare it to the observed one with a data point for each exposure.

\begin{figure}
    \centering
    \includegraphics[width=0.95\linewidth]{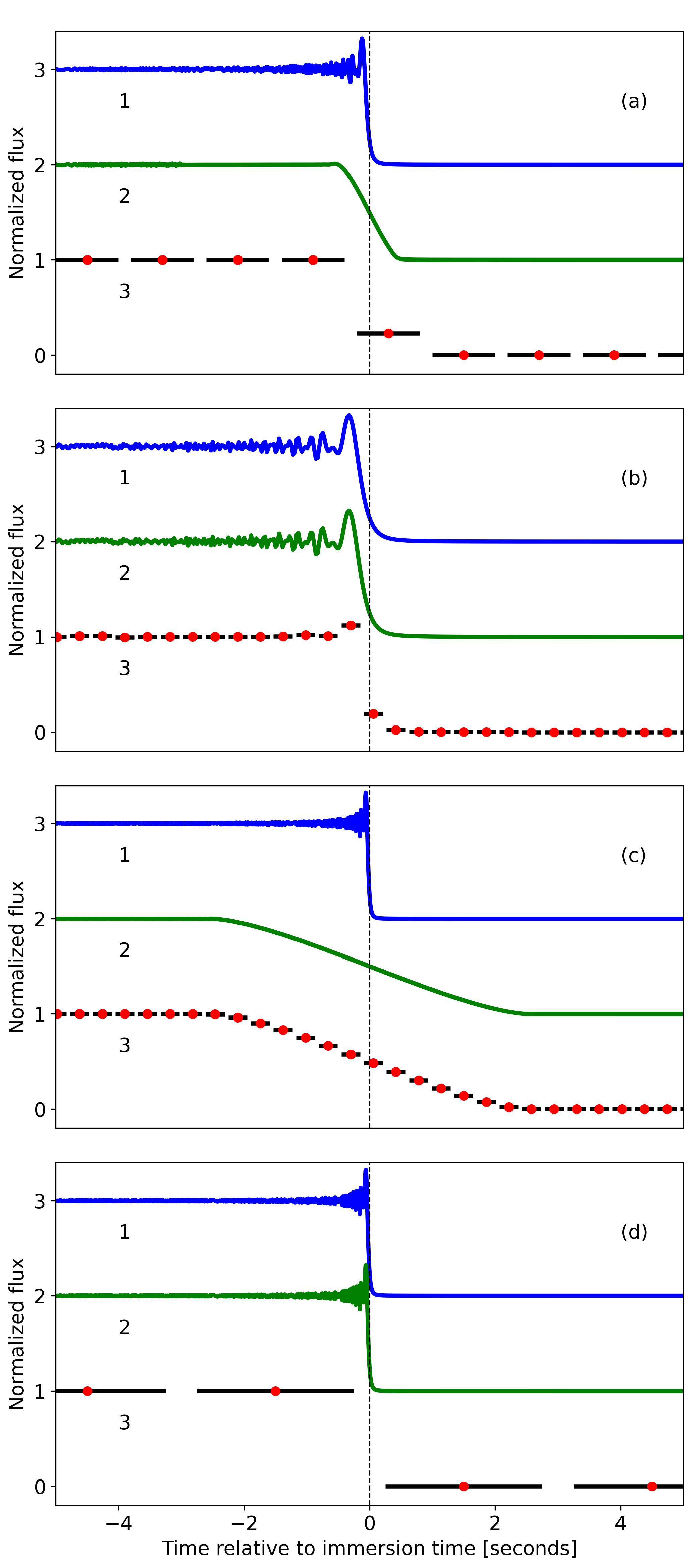}
    \caption{Modelled light curves for an occultation similar to \autoref{fig:diffraction} ($R_\circ = 50\ \textrm{km}$, $\lambda = 700\ \textrm{nm}$) for a central chord with a velocity of disappearance $V_N = 10\ \textrm{km/s}$. In each panel, (1) shows the light curve convolved with the Fresnel diffraction for an observation with a bandwidth $\Delta\lambda = 300\ \textrm{nm}$. (2) shows the model (1) convolved with the apparent size of the star. (3) shows the model (2) convolved with the exposure time, where the black lines and red dots represent the exposure duration and the final modelled light curve, respectively. The space between exposures is due to the readout time, where no data is acquired. Each panel was chosen to emphasise a feature: (a) is a base model with $D_\circ=20\ \textrm{au}$, $r_\star=5.0\ \textrm{km}$, $t_{exp}=1.0\ \textrm{s}$; (b) emphasises the Fresnel diffraction, with $D_\circ=150\ \textrm{au}$, $r_\star=0.05\ \textrm{km}$, $t_{exp}=0.3\ s$; (c) emphasises the stellar radius, with with $D_\circ=5\ \textrm{au}$, $r_\star=25.0\ \textrm{km}$, $t_{exp}=0.3\ \textrm{s}$; and (d) emphasises the instrumental response, with $D_\circ=5\ \textrm{au}$, $r_\star=0.05\ \textrm{km}$, $t_{exp}=2.5\ \textrm{s}$. We highlight that in (d) the immersion time was during the readout time, where no information were acquired.}
    \label{fig:convolved_model}
\end{figure}

\autoref{fig:convolved_model} shows the light curve model considering: (1) only the Fresnel diffraction convolved with the observation bandwidth (blue); (2) the model (1) convolved with the size of the star (green); and (3) the model (2) convolved with the exposure time (red). Panel (a) shows the models for a typical observation. Panels (b), (c), and (d) show the models when the effects of a specific parameter are the most prominent one: the Fresnel diffraction (b), the star's size (c), and the exposure time (d). Furthermore, the readout time (dead time) should also be considered since, during this period, no light flux is measured, as shown in panel (d). The mean time interval between the start of two consecutive exposures, i.e., the addition between the exposure time and readout time, is called cycle time.

The fitting procedure should provide the times of disappearance and reappearance with sub-exposure accuracy. Suppose a curve where an intermediary point between \textit{baseflux} and \textit{bottomflux} is detected, as shown in \autoref{fig:convolved_model}, panel (a), model (3). If the occultation had occurred at the start of the exposure, the relative flux of the intermediary point would be equal to the \textit{bottomflux}. If it occurred at the end of the exposure, it would be equal to the \textit{baseflux}. Then, the fraction of exposure in which the occultation happened is proportional to the fraction of relative flux drop ($\Delta f = baseflux - bottomflux$). The uncertainty of the occultation time ($\sigma_t$) can then be roughly estimated by:

\begin{equation}\label{Eq:occtime_sigma}
    \sigma_t = \frac{\sigma_{LC}}{\Delta f}t_{exp},
\end{equation}
where $\sigma_{LC}$ is the flux uncertainty.

We may notice \autoref{Eq:occtime_sigma} is not dependent on the intermediary point flux. Hence, we can still obtain a sub-exposure accuracy when the occultation happens at the start or the end of the exposure, and the intermediary point is unclear. On the other hand, if the occultation occurs in a readout time, it is impossible to identify the instant within this interval where the drop occurred. Thus, in this scenario, the uncertainty of the occultation instant is increased by the readout time.

\subsection{Stellar occultation geometry}
\label{Subsec:science-geometry}

Usually, an occultation is observed from many stations where each observation will reveal a small cross-track region of the body (as can be seen in \autoref{fig:diffraction_wavefront}). Furthermore, as shown in \autoref{Subsec:science-lightcurve}, the instants of immersion and emersion can be determined with sub-exposure accuracy, sometimes in the tenths of milliseconds. To make the most of this accuracy, the observations must have well-calibrated timings. The observer's location, either a ground-based telescope or a spacecraft, must also be known with high precision. Otherwise, combining all the observations from different locations may lead to incorrect results.

For a specific observer, the instants of immersion and emersion are the moments where the line-of-sight observer-star matches the projected limb of the body. Therefore, the relative position between the star and the body at these moments gives the distance of the limb to the body's centre in a plane perpendicular to the line-of-sight, the tangent plane. Thus, we can define a bi-dimensional Offset Frame ($x$, $y$) based on the ICRS relative to the body’s position with origin at the observer as:
\begin{align}\label{Eq:calc_xy}
    x &= D_\circ\sin[(\alpha_\star - \alpha_\circ)\cos\delta_\circ] \notag  ~,\\
    y &= D_\circ \sin(\delta_\star - \delta_\circ) ~,
\end{align}
where ($\alpha_\star$, $\delta_\star$) and ($\alpha_\circ$, $\delta_\circ$) are the astrometric ICRS coordinates of the star and body, respectively, at the occultation instant centred at the observer.

\autoref{Eq:calc_xy} requires the coordinate of the body to be known. However, in the Gaia era, the coordinate of the body presents larger uncertainty than the star's. To overcome this problem, \autoref{Eq:calc_xy} is redefined with ($\alpha_\circ$, $\delta_\circ$) being the numerically calculated position, i.e. the ephemeris. Therefore, the displacement of the observed body centre on the tangent plane ($x_c$, $y_c$) can be interpreted as an ephemeris offset, thus improving the knowledge of the body's position (see \autoref{Subsec:science-results}).

Furthermore, the precise calculation of \autoref{Eq:calc_xy} must take into account the position of the star with parallax corrected for the location of a specific ground-based observer or spacecraft as the observer moves on the space. Usually, the geocentric star coordinate is enough for an occultation on Earth. However, for occultations involving closer stars and farther bodies, the use of topocentric coordinates may be relevant. The reason is that the line-of-sight observer-star for different observers cannot be approximated by parallel lines anymore.

For instance, considering the nearest star to the Sun, Proxima Centauri, the difference between a topocentric and geocentric star coordinate may reach $2.3\ \textrm{km}$ projected at $100\ \textrm{au}$. This difference can be significant for occultations by more distant bodies or when observing an event combining ground and space-based observatories.

Mostly, the distance of the star is not relevant and can be considered at the infinity. In such cases, the coordinates ($x$, $y$) can be represented by the difference in Orthographic Projection of the geocentric observer (\autoref{Eq:orthographic-observer}) and body (\autoref{Eq:orthographic-body}) coordinates in a plane perpendicular to the direction geocentre-star:

\begin{align}
    \xi_\lambda &=  D_\lambda\cos{\delta_\lambda} \sin(\alpha_\lambda-\alpha_{\star}) ~,  \label{Eq:orthographic-observer} \\
    \eta_\lambda &= D_\lambda [\cos{\delta_{\star}} \sin{\delta_\lambda} -\sin {\delta_{\star}} \cos{\delta_\lambda} \cos (\alpha_\lambda -\alpha_{\star})] ~, \notag \\
    \notag \\
    \xi_\circ &=  D_\circ \cos{\delta_\circ} \sin(\alpha_\circ-\alpha_{\star}) ~, \label{Eq:orthographic-body}  \\
    \eta_\circ &= D_\circ [\cos{\delta_{\star}} \sin{\delta_\circ} - \sin{\delta_{\star}} \cos{\delta_\circ} \cos(\alpha_\circ - \alpha_{\star})] ~, \notag \\
    \notag \\
    x &= \xi_\lambda - \xi_\circ ~, \label{Eq:calc_xy_approximation} \\
    y &= \eta_\lambda - \eta_\circ ~, \notag
\end{align}
where ($\xi_\lambda$, $\eta_\lambda$) and ($\xi_\circ$, $\eta_\circ$) are the orthographic projection coordinates onto the tangent plane for the observer and body, respectively. The tangent plane is defined with $\xi$ positive in the direction of growing right ascension in the ICRS and $\eta$ positive in the direction of the ICRS North Pole. $D_\lambda$, $\alpha_\lambda$, and $\delta_\lambda$ are the geocentre-observer radial distance, azimuthal angle, and polar angle, respectively, in the ICRS.

This approximation has been adopted successfully by most of the published occultations such as \cite{Sicardy2011, BragaRibas2013, BenedettiRossi2019}. However, it is important to remember that using the geocentric coordinates to represent topocentric observations should be handled carefully. Besides the situation where the star distance is relevant, a more common problem may be found if the light-time correction between geocentre and observer is relevant. Considering the Earth's radius, the light-time correction is $0.021\ \textrm{s}$. With the development of more sensitive and faster cameras, shorter exposure times are achieved. Even with larger exposures, the analysis can obtain a time resolution in the order of $0.02\ \textrm{s}$, or better, if the observations present high S/N (\autoref{Subsec:science-lightcurve}). If the event's velocity is, for example, $30\ \textrm{km/s}$, this represents an error of $600\ \textrm{m}$ on the tangent plane, a significant value for occultations by small bodies like NEAs.

\subsection{Size determination and further results}
\label{Subsec:science-results}

In \autoref{Subsec:science-geometry}, we associated the instants of immersion and emersion to the projected limb of the body. Each observation can give two points on the tangent plane. Observations at the same site or path, although important as they complement each other, will provide only one effective chord in the tangent plane.

When only one effective chord is detected on a given event, the results are minimal. Therefore, we can estimate the minimum size of the body's largest dimension since the body cannot be smaller than the chord length and obtain an astrometric position.

When two or more effective positive chords are detected, the number of points on the tangent plane increases. Hence, we can search if ellipses (or circles) fit the observed chords. The fitted parameters are the apparent semi-major axis ($a'$), apparent semi-minor axis ($b'$) or apparent oblateness ($\epsilon' =  1 - b'/a'$), the position angle\footnote{Angle measured relative to the north celestial pole, turning positive into the direction of the right ascension.} of the apparent semi-minor axis ($PA_{b'}$), and the centre positions ($x_c$, $y_c$).

Large and massive bodies are more likely to have an ellipsoidal 3D shape, which can be defined by Jacobi ellipsoids or Maclaurin spheroids if the body is in hydrostatic equilibrium \citep{Chandrasekhar1987,Tancredi2008}. Thus, the body's mean apparent shape can be considered an ellipse even when many positive chords are obtained.

Nevertheless, the uncertainty of the chord extremities must be handled with care. As explained in \autoref{Subsec:science-lightcurve}, we can obtain sub-exposure uncertainties that, depending on the shadow velocity, can represent a sub-kilometric accuracy of the local limb. Suppose that the body has topographical features more prominent than the accuracy of the chord. In that case, the elliptical shape, that represents the mean surface, will be biased by the higher weight of these points, regardless if it is in a surface depression or elevation, as argued by \cite{Taylor1972}.

\begin{figure*}
\center
\includegraphics[width=\linewidth]{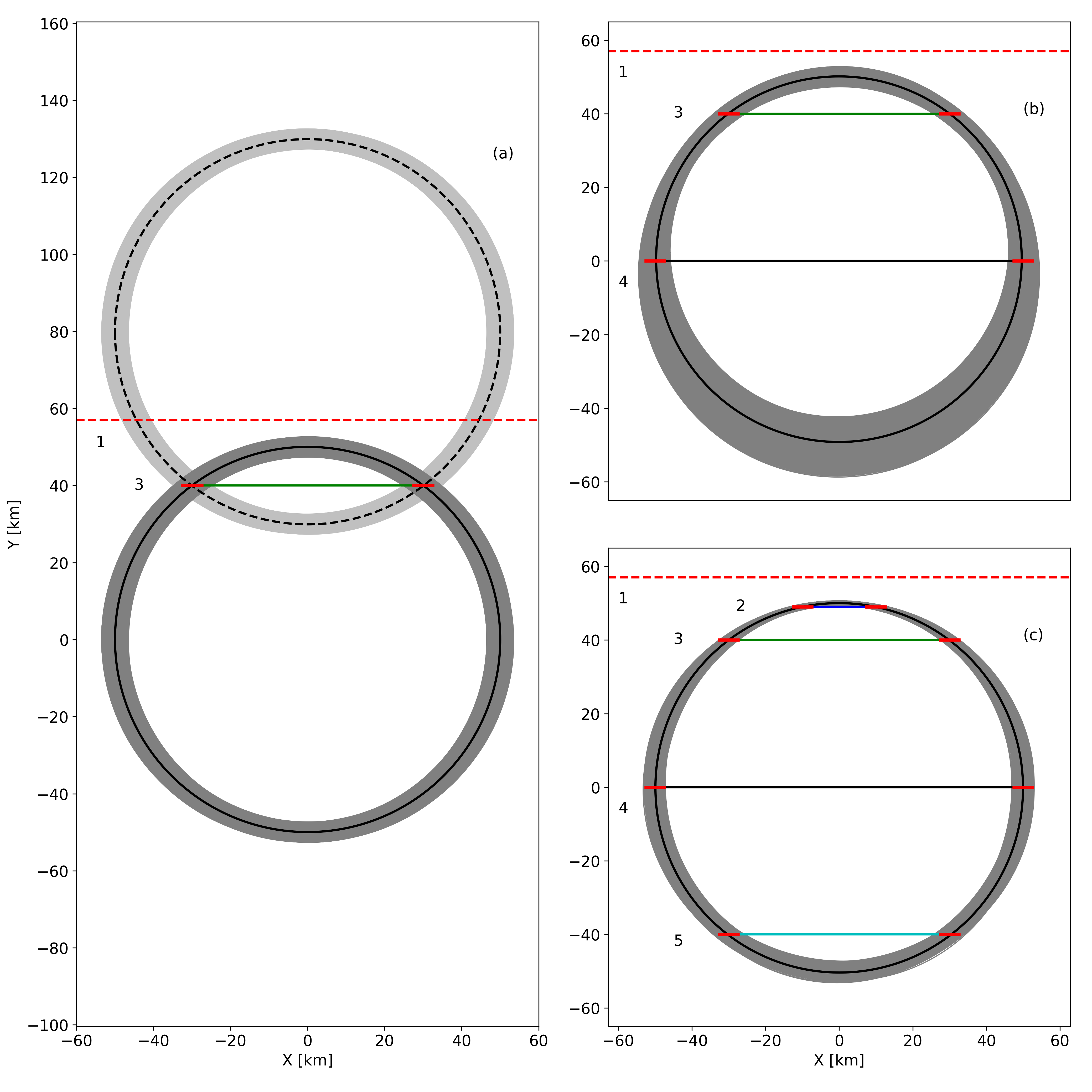}
\caption{Examples of shape fitting using one chord (a), two chords (b), and four chords (c). In (a), a circular shape with a fixed radius was adopted, and the centre was fitted to chord (3). In this case, two solutions were obtained for the centre of the circle. By considering the negative chord (1), we can eliminate the northern solutions. In (b), we fitted a circular shape to the chords (3) and (4), varying the radius and the centre position. And in (c), all five parameters of the ellipse were fitted using all the chords. The black line shows the best shape calculated, i.e., the one with the smallest global $\chi^2$. The grey region shows all the circles (or ellipses) that fit the chords within the marginal $1\sigma$ error bar. \label{fig:fit_ellipse}}
\end{figure*}

In \autoref{fig:fit_ellipse} we show shape models fitted to simulated chords. We emulate the situation in panel (a), where only one positive chord was observed fitting a circle with a fixed radius. Thus, two possible solutions for the centre position are obtained. The dashed straight line shows the negative chord. The presence of this chord removes the ambiguity in the solutions.

In panel (b) of \autoref{fig:fit_ellipse}, we fit a circle using two chords fitting for the radius and centre position. And in panel (c), we fit all the five parameters of an ellipse (apparent semi-major axis, apparent oblateness, position angle of the apparent semi-minor axis, and centre positions) to four chords. The marginal $1\sigma$ error bar, shown by the grey regions, reflects the quality of the observations, the number of chords, and the number of fitted parameters.

From the best elliptical model, we can look at the residuals in the radial direction to obtain clues about potential topographical features. For example, if a chord is located in a crater or mountain, we expect a large residual. Furthermore, if there are close chords, they should have similar values, further suggesting a topography.

Still, we must be aware of possible time issues of the observations that should not be mistaken by topographical features. As explained in \autoref{Subsec:science-geometry}, the observer must have the times inserted in the observations synchronised with a reliable time source. However, sometimes this is not possible.

In \cite{GomesJunior2015}, one of the chords was displaced by more than one minute relative to the others. The authors chose not to use this chord in the limb fitting because the remaining chords were enough for this calculation. In \cite{Ortiz2020}, the authors fitted a line to the centre of the chords and identified the best time shift for the chords with unreliable times. This method assumes that the body's apparent shape is an ellipse. In this scenario, the chords' centres should lie in a straight line connecting them.

A significant result from stellar occultations is the astrometric position. As shown above, the fitting process can determine the centre of figure on the tangent plane ($x_c, y_c$), which is the ephemeris offset at the occultation epoch. The uncertainty of the astrometric position is then calculated by combining the uncertainty of the star's position at the occultation epoch, accounting for the error in the definition of the tangent plane frame, and the uncertainty of centre determination.

This is true even when we have only one positive chord. With further considerations, we can also estimate an astrometric position in such cases. For example, \cite{Rommel2020} uses the equivalent radius of Trans-Neptunian Objects, known from previous occultations or evaluated by other sources like the radiometric technique, to fit a circle to the points, determining only the centre. \cite{GomesJunior2019} fits the single-chord to the 3D shape model of Phoebe, calculating the ephemeris offset.

In these scenarios, if the body's size is larger than the chord length, there are two possible solutions - one where the centre of the body is south of the chord and another at the north. If a close negative chord is detected, it would be possible to determine the correct position unequivocally, as can be seen in \autoref{fig:fit_ellipse} (a).

The body's apparent shape and position are direct results of occultations. But combining these parameters with external data can provide or improve the body's physical parameters. For example, \cite{Ortiz2017} and \cite{BenedettiRossi2019} obtained rotational light curves near the epoch of occultation and identified the event occurred close to the absolute brightness minimum. Thereby, they were able to determine the body's tri-axial dimensions. In addition, \cite{Morgado2021} combined the apparent shape from different stellar occultations by the Centaur Chariklo to determine that the body's shape is a triaxial ellipsoid.

Moreover, if the body's absolute magnitude is known, its geometric albedo can be calculated, as shown by \cite{Sicardy2011}. In the same sense, we can estimate the body's density if the mass is known \citep{Souami2020}. However, if the mass is not known but the rotational period is, the interval of possible densities can be calculated assuming the body is in hydrostatic equilibrium as described by the Maclaurin spheroid or Jacobi ellipsoid \citep{Tancredi2008,BragaRibas2013}.

Although occultation observations are focused on detecting positive chords, the negative chords are also of the utmost importance. When close to the limb, it will help constrain the body's size and shape. In situations where only one positive chord is detected, a close negative one can help to remove the ambiguity in the astrometric solution. Furthermore, the lack of features in the light curves is also essential information that will help on the study of the object's environment (i.e. presence of rings, dust, satellite, etc.). Finally, when only negative chords are obtained, one can map the region of ephemeris uncertainty and constrain the body's position for near-future events.

\section{The SORA Library} \label{sec:sora}

The goal of \textsc{sora} is to be an organised structure with a diversity of tools where the users can create their own pipeline for the analysis of stellar occultations. Hence, we divided the package into modules where all the information and functionalities of the subjects involved in the occultation can be stored and used efficiently.

\textsc{sora} has been made public\footnote{GitHub: \url{https://github.com/riogroup/SORA}} with version 0.2. In this publication version, the package is prepared to handle standard events, where a single Solar System body, without atmosphere, occults a single star. The technique adopted is described in \autoref{sec:occultation}. Further improvements will be discussed in \autoref{sec:conclusion}.

Developed in Python, \textsc{sora} is mainly based on classes, with attributes and methods, rather than functions. With this, we can eliminate the need of many inputs since all the data that represent the body or the star, for instance, are stored in python objects. As an example, once we define a \texttt{star} object, its Right Ascension can be accessed with \texttt{star.ra}, and the astrometric position at any time calculated with \texttt{star.get\_position(time, observer="geocenter")}.

The core development of \textsc{sora} was made using \textsc{astropy} \citep{Astropy2013}. Astropy has many functionalities to handle coordinates, time, units, etc., so we focus on stellar occultation. Furthermore, the \textsc{astroquery} package \citep{Astroquery2019} is also of major importance for accessing online databases and services, such as the Gaia catalogues, CDS VizieR\footnote{Website: \url{http://vizier.u-strasbg.fr/}}, JPL Horizons\footnote{Website: \url{https://ssd.jpl.nasa.gov/horizons/app.html}}, the Minor Planet Center\footnote{Website: \url{https://www.minorplanetcenter.net/}} (MPC) and the JPL Small-Body Database Browser\footnote{Website: \url{https://ssd.jpl.nasa.gov/tools/sbdb_lookup.html}} (SBDB).

The main modules in \textsc{sora} are: body (\autoref{Subsec:sora-body}), star (\autoref{Subsec:sora-star}), observer (\autoref{Subsec:sora-observer}), prediction (\autoref{Subsec:sora-prediction}), lightcurve (\autoref{Subsec:sora-lightcurve}), and occultation (\autoref{Subsec:sora-occultation}). Other functionalities are described in \autoref{Subsec:sora-others}.

\subsection{The \textit{body} module}
\label{Subsec:sora-body}

The \texttt{sora.body} module provides functionalities to handle the Solar System body. The main element of the module is the \texttt{Body} class that can be used to represent and manipulate the information of a single Solar System body.

\texttt{Body} objects can be created using standard Python object instantiation. The physical parameters can be input by the user or automatically downloaded from the JPL Small-Body Database Browser (SBDB). Almost all physical parameters from SBDB, if available, are set as attributes of the Body object, like albedo, diameter, density, the Standard Gravitational Parameter (GM), rotational period, pole coordinates, $B-V$ and $U-B$ colours, absolute magnitude (H), phase slope (G), and the SPK-ID (parameter used to identify the body in SPICE kernels). In addition, their 1-sigma error bars and reference source are also stored.

Other parameters provided by SBDB like the orbital-class and the spectral type in the Tholen classification \citep{Tholen1984} and SMASSII classification \citep{Bus1999} are also stored as they may be of interest for the user. We further complement the information of spectral features, given the spectral classification, as defined by \cite{Cellino2002}.

It is important to note that the SBDB does not provide information for satellites and planets. To partially overcome this issue, we offer physical parameters from the literature for the main inner and outer satellites. Any information given by the user has priority over the fetched data.

An essential tool for the body is obtaining its position in space. This functionality is provided in the \texttt{sora.ephem} module containing two main Python classes: \texttt{EphemKernel} and \texttt{EphemHorizons}.

The \texttt{EphemKernel} class was designed to facilitate the calculation of the body's position from SPICE kernels, which are files containing navigation and ancillary information as defined by the NAIF/SPICE system\footnote{Website: \url{https://naif.jpl.nasa.gov/naif/index.html}} \citep{Acton1996}. For that, we use the methods of the NAIF/SPICE toolkit through the Pythonic wrapper \textsc{spiceypy} \citep{Annex2020}.

The kernels provided to \texttt{EphemKernel} must be input by the user, as they can be obtained from different sources. However, we implemented a function that can be used to download spice kernels for small bodies from the JPL\footnote{Website: \url{https://ssd.jpl.nasa.gov/?horizons_doc}}. The \texttt{spkid}, necessary for the identification of the body in the SPICE kernel, was automatically downloaded from the SBDB. Alternatively, the \texttt{EphemHorizons} class can be used to obtain the ephemeris querying JPL Horizons web service.

Furthermore, the positions provided by either class are in the ICRS. The origin of coordinate can be \textit{barycenter}, \textit{geocenter}, or any observer defined in the \texttt{sora.observer} module (\autoref{Subsec:sora-observer}).

\subsection{The \textit{star} module}
\label{Subsec:sora-star}

The information about the star can be managed with the \texttt{sora.star} module. The \texttt{Star} class was designed to store the information about a single star and allow it to be efficiently used in the occultation process.

A \texttt{Star} object can be created by querying the Gaia-DR2 or Gaia-EDR3 tables on the CDS Vizier service. The six astrometric parameters (RA, DEC, proper motions in RA and Dec, Parallax and Radial Velocity), their uncertainties, the magnitude in the G band, and the star size are stored as parameters, when available. All the remaining data is made available as metadata.

The user can also choose to replace the parallax of Gaia-DR2 with that of \cite{BailerJones2018} and correct the proper motions of Gaia-EDR3 using the methodology provided by \cite{CantatGaudin2021}. The reason for these corrections is explained in subsections \ref{Subsec:science-prediction} and \ref{Subsec:science-lightcurve}.

Furthermore, we provide functions to calculate the angular size of the star using the methodologies of \cite{vanBelle1999} and \cite{Kervella2004}. For that, the \texttt{Star} object can do a cross-match to the NOMAD catalogue \citep{Zacharias2004}, that provides astrometric and photometric data for over 1 billion stars, and download their photometric data, which includes the B-, V-, and K-broadband magnitudes required by these methods.

Finally, the star's position can be determined at any epoch for any observer. The coordinate transformation from the catalogue epoch to any epoch is implemented using the rigorous treatment provided by \cite{Butkevich2014}. Then, the position is corrected by parallax for the observer, which can be the \textit{barycenter} (no correction), \textit{geocenter}, or any observer defined in the \texttt{sora.observer} module (\autoref{Subsec:sora-observer}). The uncertainty of the coordinates is also propagated following the procedures developed by \cite{Butkevich2014}.

\subsection{The \textit{observer} module}
\label{Subsec:sora-observer}

The \texttt{sora.observer} module provides support for defining and using the location of the observer. The main components are the \texttt{Observer}, for ground-based observer, and \texttt{Spacecraft} classes.

With the \texttt{Observer} class, the geodetic coordinates (longitude, latitude, and height) can be provided by the user or fetched from the Minor Planet Center if the site is registered. The observer position is then referred to the geocentre in the ICRS following the conventions defined by the International Earth Rotation and Reference Systems Service\footnote{Website: \url{https://www.iers.org/IERS/EN/Home/home_node}} (IERS).

The \texttt{Spacecraft} class was designed to provide the location of a spacecraft in space. For that, it uses the functionalities of the \texttt{sora.ephem} module. This is done using SPICE kernels with the spacecraft's position provided by the user, or the JPL Horizons web service, if available.

\subsection{The \textit{prediction} module}
\label{Subsec:sora-prediction}

The \texttt{sora.prediction} module provides functionalities to predict stellar occultations, organise the predictions in tables and plot the occultation maps. These outputs can be used to promote an occultation in observational campaigns.

The \texttt{prediction} function uses a \texttt{Body} object, and respective ephemeris, to calculate the body's path on the sky. The origin of the coordinates can be the geocentre or a specific observer, defined by an \texttt{Observer} or \texttt{Spacecraft} object. Then, the function downloads a list of Gaia-DR2 or Gaia-EDR3 stars from VizieR, which covers the region of the body's path for a given time interval. The path can be divided into small parts and the maximum star magnitude limited to prevent download overflow.

First, the positions of the stars are propagated to the middle epoch of the search interval. Then, to quickly identify potential occultations, the positions of the stars are matched to the ephemeris points using the \texttt{match\_to\_catalog\_sky} method provided by \textsc{astropy}. In this regard, we suggest that the step between ephemeris points be small enough not to compromise the identification of potential candidates. The function then selects those whose separations star-ephemeris are smaller than the search radius. Finally, once an event is detected, the function propagates the position of the star to the epoch of the closest approach using the rigorous treatment of \cite{Butkevich2014} and calculates the occultation parameters as described in \autoref{Subsec:science-prediction}.

We define the search radius as the body's radius increased by the ephemeris 1-sigma uncertainty if defined. If the prediction is made for the geocentre, it is inferred the user searches for any occultation on Earth. Thus the Earth's radius is added to the search interval. Furthermore, the user can always provide a custom search radius.

The output of the \texttt{prediction} function is a table in the format of the \texttt{PredictionTable} class, a class based on the \textsc{astropy} \texttt{Table} class to make the access to the prediction information easier. The information saved in the table is the epoch of the closest approach; the astrometric coordinate of the star and the body at epoch; the closest approach distance; the position angle, the shadow velocity; the distance to the body; Gaia's photometry in the G-band; Earth's longitude at the sub-star point; local solar time at sub-star point; the angular separation between the star and the Moon, and between the star and the Sun; the Gaia's (DR2 or EDR3) Source ID.

The occultation map can be plotted from the \texttt{PredictionTable} object. We consider a linear movement of the shadow on the tangent plane. Then, for each instant, the Earth's view as referred to the star is calculated, and the location of shadow is matched to the Earth's surface, determining the longitude and latitude of the centre of the shadow. Finally, the map is plotted with the Earth's view at the instant of closest approach using \textsc{cartopy} \citep{Cartopy}. The shadow's path, associated with the geodetic coordinates, is plotted over the Earth.

\begin{figure}
    \centering
    \includegraphics[width=\linewidth]{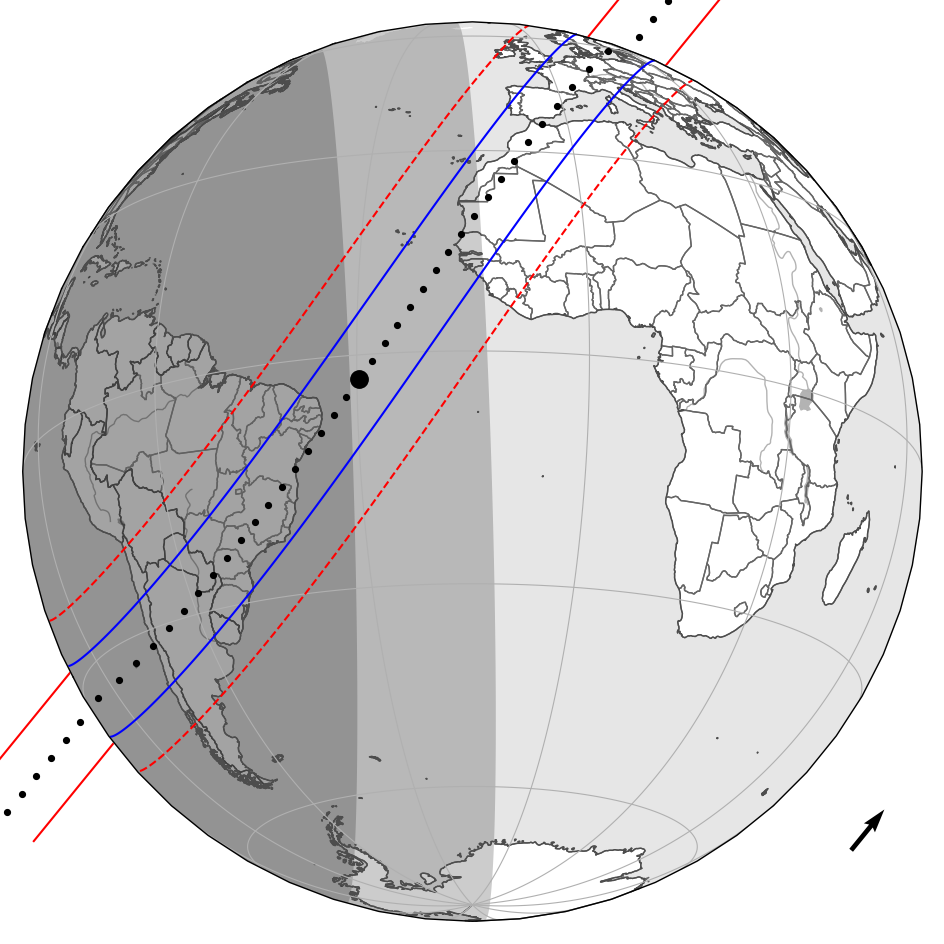}
    \caption{Occultation map showing the movement of the shadow of Quaoar over the Earth on March 27th, 2019. The arrow on the right corner gives the direction of its movement. The blue lines show the limits corresponding to the size of the shadow (the assumed body's diameter), while the red dashed ones extend these limits considering the 1-sigma error bar (here defined as $20\ \textrm{mas}$ for better visualisation). Each black dot marks the centre of the shadow separated by one minute, while the big one shows the shadow location at the instant of closest approach. The grey regions on the left are the nightshade considering the civil twilight (light grey) and the astronomical twilight (dark grey).} 
    \label{fig:occmap-example}
\end{figure}

\autoref{fig:occmap-example} shows an example of an occultation map generated by \textsc{sora}. Some of the information provided in the table can also be exhibited for completeness. Several parameters are provided to control how the map is presented.

\subsection{The \textit{lightcurve} module}
\label{Subsec:sora-lightcurve}

The \texttt{sora.lightcurve} was created to allow the user to control a light curve from a stellar occultation and determine the immersion and emersion times. The light curves are managed through the \texttt{LightCurve} class.

The functions in this module can be divided into four main procedures:
\begin{itemize}
    \item Set the parameters of the model;
    \item Prepare the light curve to be fitted;
    \item Create light curve models and fitting of the occultation parameters; 
    \item Plot and export information.
\end{itemize}

SORA does not provide any functionality to obtain the light curve from the observations. Deriving an optimised light curve from photometry demands dedicated software for this task. Several software are available, like PRAIA \citep{Assafin2011}, Tangra\footnote{\url{http://www.hristopavlov.net/Tangra3/}} \citep{Tangra2020}, and PyMovie\footnote{\url{https://occultations.org/observing/software/pymovie/}} \citep{Pymovie2019}, and the users can choose whichever software they find more suitable. Then, the light curve can be imported into a \texttt{LightCurve} object.

A \texttt{LightCurve} object is instantiated from the mid-time of the observations, flux, and exposure time provided by the user. The time scale is expected as UTC, but the user can input an \texttt{Astropy} Time object in whichever time scale that \textsc{sora} will handle it correctly. The flux uncertainty can also be input. When no flux uncertainty is provided, it is calculated as the dispersion of the light curve photometry outside the occultation drop.

As discussed in section \ref{Subsec:science-lightcurve}, five parameters have to be considered when modelling an occultation light curve: the wavelength and bandwidth of the setup (which is related to the diffraction); the exposure time; the apparent diameter of the star; the distance of the observer to the occulting body, and the velocity of star disappearance (reappearance). The bandwidth and exposure time must be provided by the user, while the apparent diameter, distance and velocity can be automatically defined by the \texttt{Occultation} class (\autoref{Subsec:sora-occultation}), once the \texttt{LightCurve} object is associated with an event.

A normalisation method is provided to reduce noises and trends caused by observational fluctuations unrelated to the occultation. A polynomial is calculated using the data outside the occultation and propagated for the whole curve. This procedure is essential mainly for the cases where no calibration star is present in the field preventing differential photometry.

To automatically identify an occultation in the light curve, we provide a function based on the box-fitting least-squares algorithm described in \cite{Kovacs2002}. Instead of a period search, we compute the statistic as it would be performed in the folded light curve (equations 1 to 4 in the referred work). In summary, we fit a step function to the data with varying positions, widths, and depths and collect the best fit as the most likely result. This function is not intended to be used as a blind estimator but rather as a faster technique to find initial parameters in occultation events. Moreover, systematic effects in the data can heavily skew the results and should always be considered; therefore, a final validation is up to the user. Finally, a more robust fitting procedure is still required to obtain precise immersion and emersion instants.

The base diffraction model of \cite{Roques1987} is normalised to \textit{baseflux} and \textit{bottomflux} as 1 and 0, respectively. Thus, we must normalise the observed light curve to these same values. Then we can compare the normalised light curve ($\phi_{obs}$) with the modelled one ($\phi_{cal}$) and compute the difference for each data point $i$. Finally, the $\chi^2$ statistics are computed using:
\begin{equation}\label{Eq:chi-square}
    \chi^2 = \sum^N_{i=1} \frac{(\phi_{i,obs}-\phi_{i,cal})^2}{\sigma_i^2} ~,
\end{equation}
where $\sigma_i$ is the photometric uncertainty of each data point provided by the user and $N$ is the total number of points considered.

The process of fitting implemented in \textsc{sora} consists in minimising \autoref{Eq:chi-square} calculating $\phi_{cal}$ using the following procedures:

\begin{enumerate}
    \item A Monte Carlo approach is used to generate instants of immersion and emersion uniformly distributed within the interval given by the user. This approach is interesting because it can test an extensive range of parameters chosen by the user, thus giving the user complete control of the process.
    
    \item For each pair of immersion and emersion times, we simulate a light curve for a bar shape occultation \citep{Roques1987} considering the normal velocity, the light diffraction, the size of the star, and the exposure time, as described in \autoref{Subsec:science-lightcurve}.
    
    \item The simulated light curve is compared to the observed one, and the chi-squared statistics are computed using \autoref{Eq:chi-square}.
    
    \item The best-fitted model is found for the immersion and emersion times that produces the global minimum chi-squared value \citep[Chapter 15.1]{NumericalRecipes}.
    
    \item The confidence levels for each parameter is calculated using constant $\chi^2$ boundaries in a marginal distribution. If a given estimate $\chi^2(\mathbf{a}_j)$ follows a $\chi^2_N$ distribution, then the value $\chi^2(\mathbf{a}_0)$ that minimises the error in that universe will follow a $\chi^2_{N-M}$ distribution. Therefore, it follows that the quantity $\Delta\chi^2 \equiv \chi^2(\mathbf{a}_j) - \chi^2(\mathbf{a}_0)$ will follow a $\chi^2$ distribution with $M$ degrees of freedom (see \cite{NumericalRecipes}, Chapter 15.6, Theorem C). By default, \textsc{sora} considers the $1\sigma$ one-dimensional marginal error, in which case $\Delta\chi^2 = 1$, whatever is the number of data points. In other words, a value $K$ draw from the $\Delta\chi^2_{M = 1}$ distribution such that the probability $P(\Delta\chi^2_{M=1} < K) = 0.6834$ (a table with values up to $M \le 6$ is given in the aforementioned work).
\end{enumerate}

It is important to note the immersion and emersion times can be fitted independently (considering only a small part of the light curve) or jointly. To avoid significant computational processing, the user can set the number of simulated light curves balanced by the region used in the fit. Data points far from the immersion and emersion instants will not significantly affect the result.

Moreover, the user can fit other parameters of the light curve model, like the \textit{baseflux}, \textit{bottomflux}, the velocity of the occultation normal to the surface, the apparent diameter of the star, etc., by external iteration combining the resulting \texttt{ChiSquare} objects (details in \autoref{Subsec:sora-others}).

The \texttt{LightCurve} class also has some visualisation methods to help the user to access the data and the models easily. Plots like \autoref{fig:convolved_model} can be obtained.

\subsection{The \textit{occultation} module}
\label{Subsec:sora-occultation}

The \texttt{sora.occultation} module is the main module of the library. The functionalities presented here are focused on the management and processing of the occultation information to obtain the results. The main feature is the \texttt{Occultation} class that organises the python objects created with the classes described in the previous subsections.

An \texttt{Occultation} instance is created with a \texttt{Star} and a \texttt{Body} objects, the instant of occultation and a location reference, usually the geocentre. Thus, it calculates the occultation parameters necessary for the light curve model and occultation map.

Then, the user can add the chords to the \texttt{Occultation} object, providing an \texttt{Observer} or \texttt{Spacecraft} object, and a \texttt{LightCurve} object. The velocity of the shadow, distance to the body and the star's apparent size are passed to the \texttt{LightCurve} if the occultation instants were not fitted before being added to the \texttt{Occultation}.

A \texttt{Chord} class was developed to link the \texttt{Occultation} and individual observation. This auxiliary class organises the information for a single chord. It is not defined by the user but can be accessed.

We have implemented in the \texttt{Chord} class the ability to project the chord on the tangent plane using the precise methodology defined by \autoref{Eq:calc_xy} and the approximated one defined by \autoref{Eq:calc_xy_approximation}. The accurate method calculates the position on the plane using the star, the body, and the observer for each given instant. Thereby, we prepare \textsc{sora} for circumstances that may be rare nowadays, as will be discussed in \autoref{sec:conclusion}.

A function is also provided to fit a straight line to the centre of the positive chords and determine the best time shift for each chord. As explained in \autoref{Subsec:science-results}, for an elliptical shape, the centre of the chords should lie in a straight line connecting them. However, we call the attention that this procedure should be used with care, as it can mask topographic features.

Finally, with the \texttt{Occultation} object, we can fit the body's mean surface to the chords' extremities using an elliptical shape. The function determines the five parameters of the ellipse: the apparent semi-major axis, the apparent oblateness, the position angle of the apparent semi-minor axis, and the centre position following these procedures:

\begin{enumerate}
    \item A Monte Carlo approach is used to create a uniform distribution for each of the parameters described above within the intervals given by the user. Here, the Monte Carlo approach is even more critical, as in many cases, a large number of shapes could fit the data. Therefore, in this step, the user should control the exact range of parameters to be tested. That allows, for instance, to filter the ellipses based on external information, such as a predetermined size or shape.
    
    \item For each set of parameters, an ellipse will be generated using standard geometrical equations. Oblateness equals zero defines a circle, therefore the position angle will degenerate.
    
    \item The radial difference between the chord's extremities to the ellipse is calculated, and the $\chi^2$ is computed using \autoref{Eq:chi-square_ellipse}.

\begin{equation}\label{Eq:chi-square_ellipse}
    \chi^2 = \sum^N_{i=1} \frac{(r_{i,obs}-r_{i,cal})^2}{\sigma_i^2 + \sigma_{model}^2} ~,
\end{equation}
where $(r_{i,obs} - r_{i,cal})$ is the difference between each chord extremity $i$ and the tested shape limb and $\sigma_i$ is the uncertainty of the $i$-th chord extremity.

In \textsc{sora}, the current implementation calculates $(r_{i,obs} - r_{i,cal})$ as the difference between each chord extremity "i" and the calculated limb in the radial direction, concerning the centre of the ellipse, as shown in \autoref{fig:fit_ellipse_radial}. The uncertainty of the chord is mostly in the chord direction caused by timings, disregarding any error in the observer's position. Thus, the uncertainty is projected in the radial direction to account for the change in parameterisation. This approach simplifies the calculation making it faster, and matches very well the direction of $\sigma_{model}$, explained below.

\begin{figure*}
\center
\subfigure{\includegraphics[width=.49\linewidth]{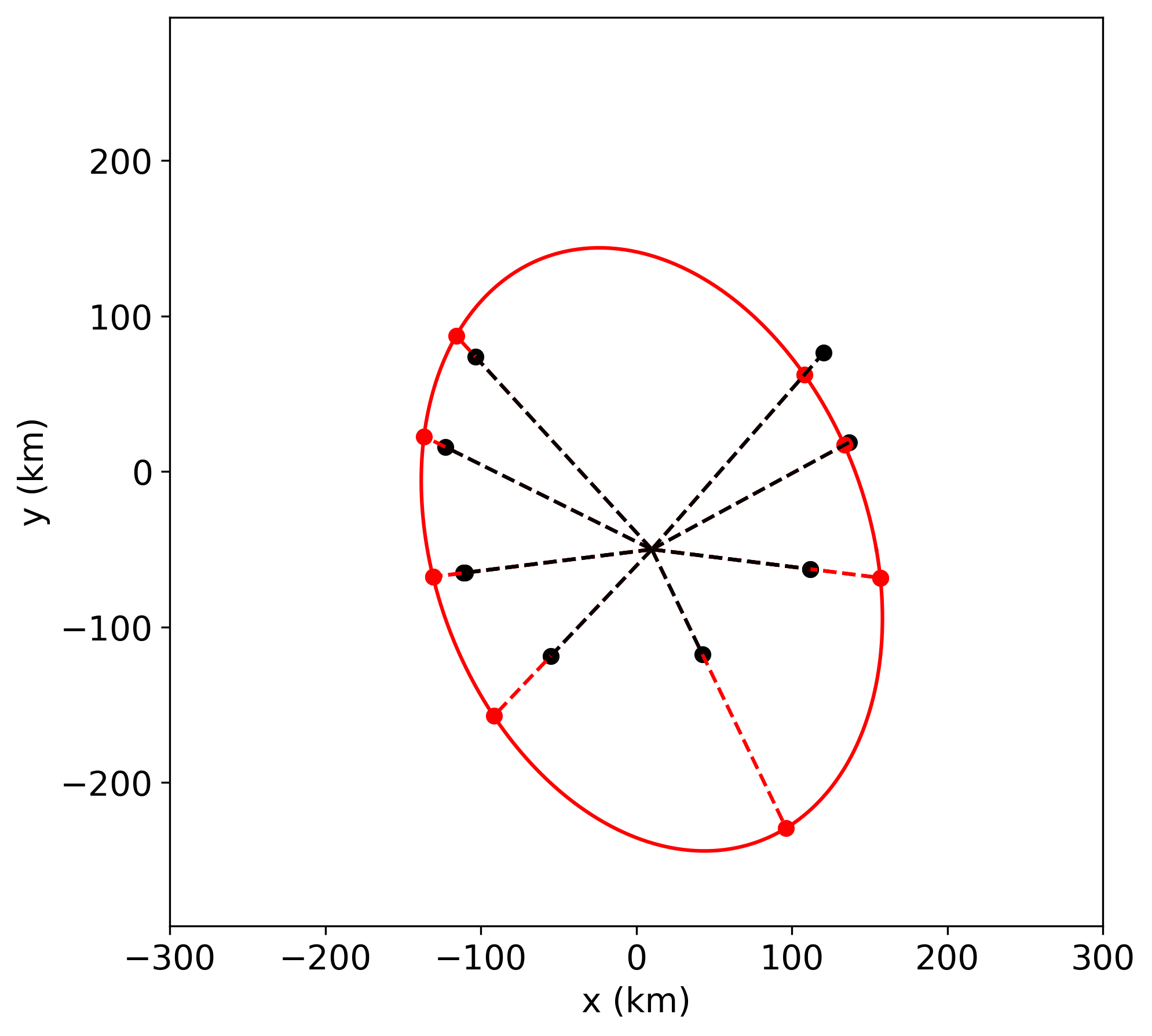}
}
\subfigure{\includegraphics[width=.49\linewidth]{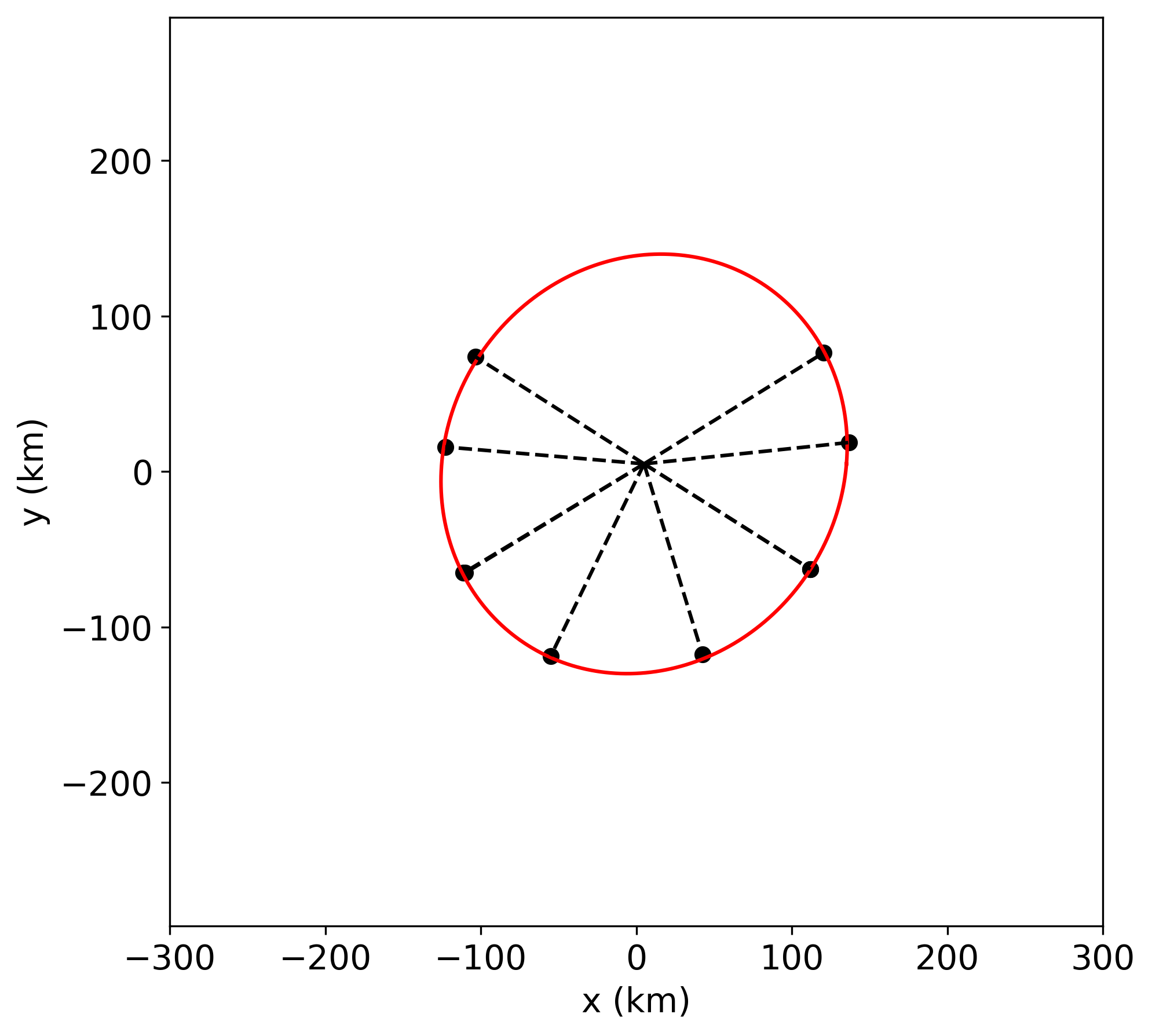} }
\caption{Figure showing the current fitting method implemented in \textsc{sora}. In the left panel, we show the chord extremities of the chords as black points on the tangent plane. Then, for each ellipse tested, we compute the radial difference between these points and the elliptical limb (red dashed lines), accounted from the ellipse's centre. Next, the uncertainty of each chord extremity is projected in the radial direction. Finally, the $\chi^2$ is computed using \autoref{Eq:chi-square_ellipse}, and the best ellipse is found as the global minimum $\chi^2$ (right panel). It is up to the user to provide a suitable interval of search.} \label{fig:fit_ellipse_radial}
\end{figure*}

Contrary to expectations, the chords are not an exact parallel among themselves due to the variation in the relative velocity between observers in different latitudes due to Earth's rotation. At the same time, the two extremities of the same chord may present different directions, especially for slow events, as the movement can not be represented by a linear motion anymore (see \autoref{Subsec:science-prediction}). Additional features such as fitting based on the direction of the chord are planned for future versions.

The last parameter in \autoref{Eq:chi-square_ellipse} is the $\sigma_{model}$. In \autoref{Subsec:science-results} we showed that, in some cases, the position of the chord on the tangent plane might be too precise, with uncertainties in the sub-kilometre level, thus smaller than possible topographic features. In such cases, the elliptical model, which accounts for the mean surface, is not appropriate anymore. Using these points to fit an ellipse could cause a bias in the result due to their higher weight. To overcome this issue, \cite{Morgado2021} included an uncertainty to the elliptical model ($\sigma_{model}$), accounting for the difference between the elliptical model and the real shape. Usually, $\sigma_{model}$ can be thought of as a typical dimension of the expected topography, accounted in the radial direction. Note that, for instance, topographic features in the 10 km level were observed in Uranus's five major satellites \citep{Schenk2020}, and up to 13 km for the Plutino (208996)  $2003 AZ_{84}$ \citep{DiasOliveira2017}. $\sigma_{model}$ is a parameter the user can provide to decrease the weight of the chords, thus obtaining an adequate mean elliptical model or completely ignoring it. By default, \textsc{sora} sets this value as zero.

    \item Finally, the $\chi^2$ marginal distribution is determined using all ellipses, and the best fit is found for the ellipse where the chi-squared value is the global minimum \citep[Chapter 15.1]{NumericalRecipes}.
    
    \item The error bar for each parameter is determined by the criterion $\chi^2 < \chi^2_ {min} + \Delta\chi^2$. By default, \textsc{sora} considers the $1\sigma$ marginal error bar where $\Delta\chi^2$ equals to $1$. However, the user have control and can choose the intended confidence interval ($\Delta\chi^2$) \citep[Chapter 15.6]{NumericalRecipes}. \autoref{fig:chi_square} shows the chi-square distribution of one parameter fitting using the method here presented. Note the parabolic shape expected from the chi-square distribution.
    
\begin{figure}
\center
\includegraphics[width=\linewidth]{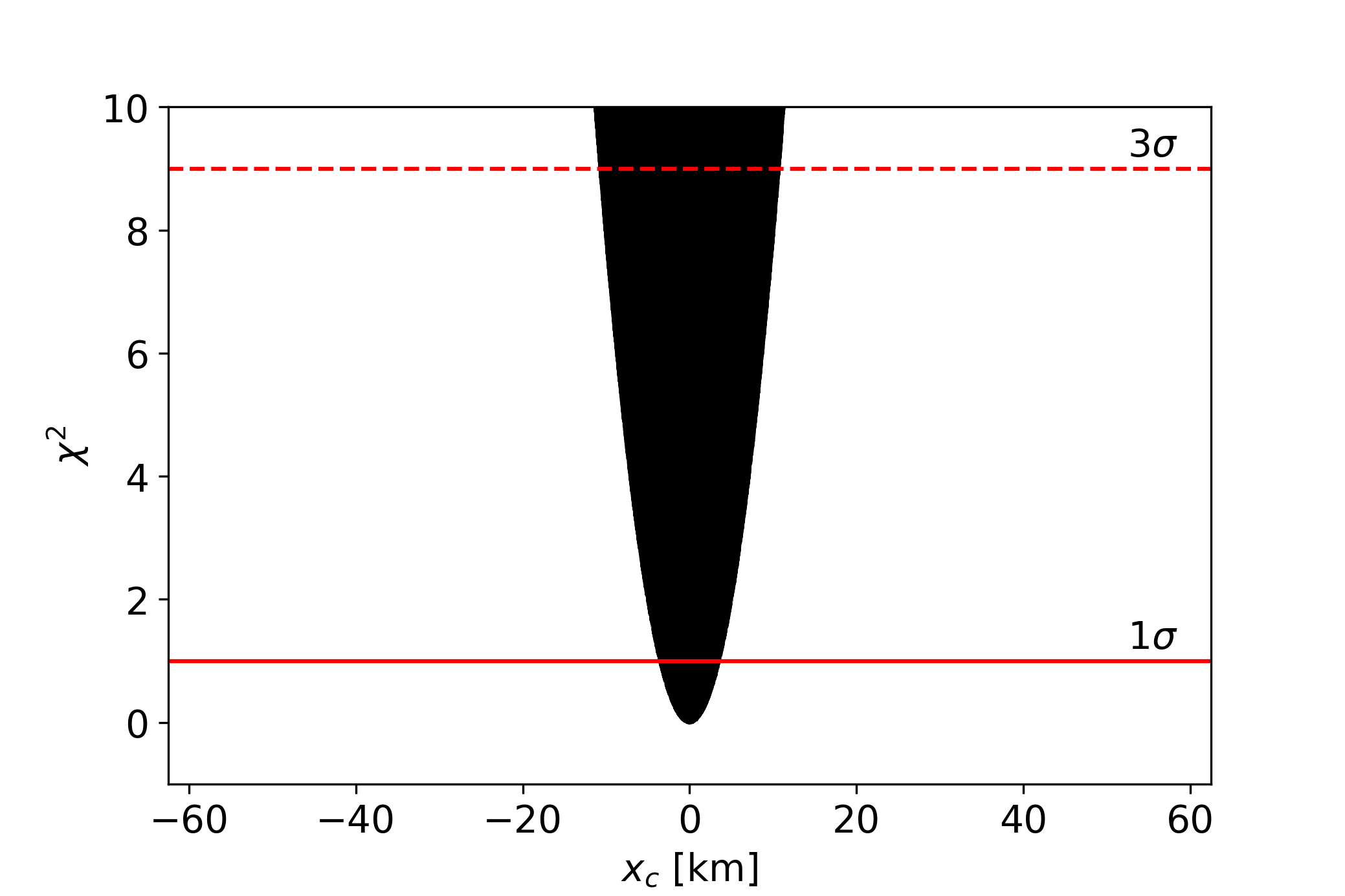}
\caption{Chi-square distribution obtained from the fit presented in \autoref{fig:fit_ellipse} for the centre position in $x$ using the method presented in \autoref{Subsec:sora-occultation}. The horizontal red lines show the limits within the marginal $1\sigma$ (solid) and $3\sigma$ (dashed) error bars. Note the parabolic shape expected from the chi-square distribution. \label{fig:chi_square}}
\end{figure}
\end{enumerate}

\textsc{sora} allows for the user to fit the ellipse using all the chords added to the \texttt{Occultation} object, a fraction of them, or even ignoring chord extremities. The default number of ellipses tested is 10 million, usually enough for an appropriate range of searches. However, the user can set this value or combine different calls to the fitting function.

Furthermore, a function that considers the path of the negative chords on the tangent plane and discards the ellipsis that crosses it is available. This functionality is essential when a negative chord is observed close to the body's limb, thus constraining the shape model. In cases where the observation has a significant readout time, the user can set the function not to filter the ellipses that cross the path of the negative chord within this interval since the observation would not detect them.

When the best ellipse is found, the shadow velocity normal to the local surface can be obtained. Then, the user can fit the occultation instants again if necessary.

Observers in different locations will usually detect the occultation in different moments. To obtain an astrometric position of the object from the occultation event, combining all these observations, the ephemeris offset is supposed to be constant during the time interval of the observations. From the ellipse fitting, the obtained centre position on the tangent plane is the ephemeris offset at the body's distance. Therefore, the astrometric position is calculated by applying this offset to the ephemeris position. The user can choose for which instant and location the astrometric position is provided. However, it should not be far from the locations and times of the observations.

As mentioned, the object size and shape determined by the stellar occultation can improve other physical parameters. However, this depends on external knowledge. For example, the absolute magnitude is needed to calculate the geometric albedo ($p_V$). SORA has a function that calculates the geometric albedo based on the user's input,  shown in \autoref{Eq:albedo}.
\begin{equation}
    p_V = 10^{H_\odot - H_\circ}\times\left(\frac{149597870.7}{R_{eq}}\right)^2
    \label{Eq:albedo}
\end{equation}
where $R_{eq}$ is the radius of the circle which area is equivalent to the fitted ellipse ($R_{eq} = \sqrt{a \times b}$), measured in kilometres, $H_\odot$ is the absolute magnitude of the Sun, whose default value is $H_\odot=-26.74$, and $H_{\circ}$ is the absolute magnitude for the object.

The final output of the \texttt{Occultation} object is the post-diction map. Here we provide the occultation map automatically updated with the occultation information and results. The observer locations are shown, identifying the positive or negative ones. The shadow's size and path over the Earth are corrected with the calculated parameters.

\subsection{Other functionalities}
\label{Subsec:sora-others}

\textsc{sora} also contains other functionalities to help the user analyse the occultation. The \texttt{sora.extra} module was created to collect classes and functions that are not directly associated with the occultation subjects.

The output of the light curve and ellipse fittings are python objects created from the \texttt{ChiSquare} class. This class was developed to be a data table where the chi-squared statistics are associated with the parameters tested. We can obtain the parameters for the minimum $\chi^2$, the marginal uncertainty, or plot the data. Multiple \texttt{ChiSquare} objects, resulting from different runs of the fitting functions, can be combined into one to be analysed jointly.

The SBDB service provides the physical parameters of the body with a given unit, uncertainty, reference, and some notes. To better organise and use these parameters, we developed the \texttt{PhysicalData} class based on the \textsc{astropy}'s \texttt{Quantity} class. The main information is the value and unit of the physical parameter. The uncertainty, reference, and notes are stored as attributes of the \texttt{PhysicalData} object and can be accessed if needed.

Some of the classes in \textsc{sora} have functions to plot their information and help the user to visualize the data, like the \texttt{ChiSquare}, \texttt{Chord} and \texttt{LightCurve} classes. There are also plot functions in the \texttt{sora.extra} module, like the one that draws ellipses on the tangent plane. These plots were developed using \textsc{matplotlib} \citep{Matplotlib}.

An important note about \textsc{sora} visualisation functions is that they do not save or plot the information on-screen themselves. Because of this, the user can combine different plot functions or use the \textsc{matplolib} tools to improve the visualisation. The only exception to this rule is the function that plots the occultation maps where the user controls the visualisation through input parameters.


\section{Summary and Future Work}\label{sec:conclusion}

\textsc{sora} is a python library for the reduction and analysis of stellar occultations. The goal is to provide functionalities related to stellar occultations where the users can develop their own pipeline. 

In the current version of the package, v0.2, the implemented tasks focus on analysing stellar occultations involving a single star and a single body, which is usually the case. The methodology adopted for this is presented in \autoref{sec:occultation}.

\textsc{sora} is intended to be as precise as possible. First, the star and the body positions are computed for an observer accurately without assumptions. Then, to combine the chords, we assume a constant offset between the body ephemeris and the star position.

The package is divided into modules organising the tools related to each occultation subject: body, star, observer, light curve, and a module that concatenates these tools for the occultation itself. In \autoref{sec:sora}, we described all the modules in \textsc{sora}, their goals and usage.

For each of these modules, a main python class was developed and can be used to define a single element with its attributes. For example, the \texttt{Star} class of the \texttt{sora.star} module is used to define a single star with its information, like coordinates. Using classes, with attributes and methods, instead of functions, we reduce the number of inputs since most can be downloaded from web services and stored in python objects.

\textsc{sora} is documented online and can be accessed at \url{https://sora.readthedocs.io/}. We include how to install and start using \textsc{sora}, detailed examples, and guidelines. \textsc{sora} is under open-source license and can be accessed on \url{https://github.com/riogroup/SORA}. The users can also participate in the development of \textsc{sora} using the GitHub procedures.

We have validated each specific step, comparing the functions and methods within \textsc{sora} with well-established astronomical software like the SPICE/NAIF toolkit and the SOFA library. We also have compared the timings obtained between \textsc{sora} and \textsc{occult} for an unpublished occultation. The times and respective uncertainties obtained are comparable within their 1$\sigma$ error bar.

Different methods for the reduction of stellar occultations can be implemented in the future. For instance, \cite{Leiva2020} and \cite{Strauss2021} adopt a Bayesian approach. They model the occultation using different priors for the body and star. They then compare the modelled light curves with the observed ones, skipping the necessity of determining the instants of immersion and emersion.

Furthermore, with the increase in the number of stellar occultations observed, events involving bodies with atmospheres, rings, satellites, binaries, etc., and even occultations by multiple stars, may become more common. Functionalities to analyse occultations in such scenarios are already being planned.

Some minor effects are not detectable by current observations. However, with the improvements in catalogues, ephemeris, equipment, and setups, they may be noticeable soon. For instance, as already discussed in \autoref{Subsec:science-geometry}, we make it possible to analyse occultations where the distance of the star is relevant or to consider the light-time correction between the geocentre and observer on observations with high-time resolution.

Further improvements are foreseen, such as correcting the relative position between body and star due to the gravitational deflection caused by the occulting body. For example, the gravitation deflection caused by Pluto of a light ray close to the body's surface as observed on Earth is about $200\ \textrm{m}$. Furthermore, we can extend this analysis to the deflection caused by the Sun and planets, which are significant in the sub-mas regime \citep{Klioner2003}.

When the velocity of the occultation is small, combining observations may require constraints on the rotational parameters. An example is the occultation of a Gaia-EDR3 star (source id 2568323823770876672) by Mars on September 10th, 2020. The event's velocity was $0.12\ \textrm{km/s}$, slower than Earth's rotation of $0.5\ \textrm{km/s}$, making the shadow last about 14 hours over the Earth. 

Suppose a body with a triaxial ellipsoidal shape and rotation period of $7.6\ \textrm{h}$, a typical value for TNOs \citep{Thirouin2010}, in the same circumstance. The body could rotate almost twice during the event, and different observers on the same path could detect chords with different lengths. If the body has a diameter of $1000\ \textrm{km}$, the difference between the immersion and emersion times for a central chord would be $\sim 2\ \textrm{h}$, so we should consider the rotation over all the observations. To overcome this problem, we intend to implement the use of rotational elements in \textsc{sora} in the analysis of the occultation.

Complementary to this approach is the analysis of multiple occultations. For example, \cite{Morgado2021} analysed 11 occultations by Chariklo between 2013 and 2020, constrained by the rotational elements, to fit a triaxial ellipsoidal shape. \textsc{sora} will adopt a similar procedure. The user will choose what parameters will be constrained when fitting multiple occultations: rotational period, shape, astrometry, pole, etc.

Another essential feature to be included in \textsc{sora} is the ability to handle 3D shape models. For example, \cite{GomesJunior2019} fitted the occultation chords to the 3D shape model of Phoebe. Thus, they could determine the sub-observer longitude at the occultation epoch and improve the rotational period. In contrast, the user can obtain the projection of the timings on the tangent plane with \textsc{sora} and use them on software that combine the occultation chords with other kinds of observations, like 3D models from rotational light curve inversion.

In conclusion, stellar occultations can be used to obtain many physical parameters of the occulting body and its vicinity with high accuracy. SORA was developed to be a precise, efficient, fast, and easy to use tool to handle stellar occultation data. Many more valuable features are planned and will be included in the library in due time.

\section*{Acknowledgements}\label{acknowledgements}
We would like to thank Luana Liberato, Giuliano Margoti, Chrystian L. Perreira, and the members of the Rio/Paris/Granada collaboration for their time to perform tests and validations which improved the quality of SORA and this manuscript.
This work was carried out within the “Lucky Star" umbrella that agglomerates the efforts of the Paris, Granada and Rio teams, which is funded by the European Research Council under the European Community’s H2020 (ERC Grant Agreement No. 669416).
This work has made use of data from the European Space Agency (ESA) mission Gaia (\url{https://www.cosmos.esa.int/gaia}), processed by the Gaia Data Processing and Analysis Consortium (DPAC, \url{https://www.cosmos.esa.int/web/gaia/dpac/consortium}).
Part of this research was supported by INCT do e-Universo, Brazil (CNPq grant 465376/2014-2).
The following authors acknowledge the respective grants:
ARGJr FAPESP 2018/11239-8; 
BEM CNPq 150612/2020-6; 
GBR CAPES-PRINT/UNESP Process 88887.310463/2018-00, Project 88887.571156/2020-00;
RCB FAPERJ/PDR E26/202.125/2020;
FLR PROEX/CAPES Finance code 001;
MVB-H FAPERJ/PDR grant E26/200.480/2020;
FB-R CNPq grant 314772/2020-0;

\section*{Data Availability}

No new data were generated or analysed in support of this research. Instructions for downloading and installing \textsc{sora} can be found on PyPI (\url{https://pypi.org/project/sora-astro/}), on GitHub (\url{https://github.com/riogroup/SORA}) and on our online documentation (\url{https://sora.readthedocs.io/}).


\bibliographystyle{mnras}
\bibliography{main}

\bsp	
\label{lastpage}
\end{document}